\documentclass[sn-mathphys]{sn-jnl}

\usepackage{graphicx}%
\usepackage{multirow}%
\usepackage{amsmath,amssymb,amsfonts}%
\usepackage{amsthm}%
\usepackage{mathrsfs}%
\usepackage[title]{appendix}%
\usepackage{xcolor}%
\usepackage{textcomp}%
\usepackage{manyfoot}%
\usepackage{booktabs}%
\usepackage{algorithm}%
\usepackage{algorithmicx}%
\usepackage{algpseudocode}%
\usepackage{listings}%
\usepackage{natbib}
\usepackage{bm}
\usepackage{doi}
\usepackage{tabularx}
\newcommand{\indep}{\perp \!\!\!\perp}
\newcommand*\rot{\rotatebox{90}}

\theoremstyle{thmstyleone}%
\theoremstyle{thmstyletwo}%
\newtheorem{remark}{Remark}%

\theoremstyle{thmstylethree}%

\begin{document}

\title[DTDCKe]{Derandomized Truncated D-vine Copula Knockoffs with e-values to control the false discovery rate}


\author[1]{\fnm{Alejandro} \sur{Román Vásquez}}\email{arv@xanum.uam.mx}
\equalcont{These authors contributed equally to this work.}

\author[2,3]{\fnm{José Ulises} \sur{Márquez Urbina}}\email{ulises@cimat.mx}
\equalcont{These authors contributed equally to this work.}

\author*[4]{\fnm{Graciela} \sur{González Farías}}\email{farias@cimat.mx}
\equalcont{These authors contributed equally to this work.}

\author[1]{\fnm{Gabriel} \sur{Escarela}}\email{ge@xanum.uam.mx}
\equalcont{These authors contributed equally to this work.}

\affil[1]{\orgdiv{Departamento de Matemáticas}, \orgname{Universidad Autónoma Metropolitana Unidad Iztapalapa}, \orgaddress{\street{Av. San Rafael Atlixco 186, Col. Vicentina}, \city{Iztapalapa}, \postcode{09340}, \state{CDMX}, \country{Mexico}}}

\affil[2]{\orgdiv{Unidad Monterrey}, \orgname{Centro de Investigación en Matemáticas, A.C.}, \orgaddress{\street{Av. Alianza Centro 502, PIIT}, \city{Apodaca}, \postcode{66628}, \state{NL}, \country{Mexico}}} 

\affil[3]{\orgdiv{Consejo Nacional de Humanidades, Ciencia y Tecnología}, 
\orgaddress{\street{Av. Insurgentes Sur 1582, Col. Crédito Constructor}, \city{Benito Juárez}, \postcode{03940}, \state{CDMX}, \country{Mexico}}}

\affil*[4]{\orgdiv{Departamento de Probabilidad y Estadística}, \orgname{Centro de Investigación en Matemáticas, A.C.}, \orgaddress{\street{Jalisco S/N, Valenciana}, \city{Guanajuato}, \postcode{36023}, \state{GTO}, \country{Mexico}}}


\abstract{
The Model-X knockoffs is a practical methodology for
  variable selection, which stands out from other selection strategies
  since it allows for the control of the false discovery rate (FDR),
  relying on finite-sample guarantees. In this article, we propose a
  Truncated D-vine Copula Knockoffs (TDCK) algorithm for sampling
  approximate knockoffs from complex multivariate distributions. Our
  algorithm enhances and improves features of previous attempts to
  sample knockoffs under the multivariate setting, with the three main
  contributions being: 1) the truncation of the D-vine copula, which
  reduces the dependence between the original variables and their
  corresponding knockoffs, improving the statistical power; 2) the
  employment of a straightforward non-parametric formulation for
  marginal transformations, eliminating the need for a specific
  parametric family or a kernel density estimator; 3) the use of the ``rvinecopulib'' R package offers better flexibility than the existing fitting vine copula knockoff methods. To
  eliminate the randomness in distinct realizations resulting in
  different sets of selected variables, we wrap the TDCK method with
  an existing derandomizing procedure for knockoffs, leading to a
  Derandomized Truncated D-vine Copula Knockoffs with e-values
  (DTDCKe) procedure. We demonstrate the robustness of the DTDCKe
  procedure under various scenarios with extensive simulation
  studies. We further illustrate its efficacy using a gene expression
  dataset, showing it achieves a more reliable gene selection than
  other competing methods, when the findings are compared with those
  of a meta-analysis. The results indicate that our Truncated D-vine
  copula approach is robust and has superior power, representing an
  appealing approach for variable selection in different multivariate
  applications, particularly in gene expression analysis.}
\keywords{Multiple hypothesis testing, Model-X knockoffs, Vine copulas, Lasso regression, Gene expression data, High-dimensional
low sample data}



\maketitle

\section{Introduction}

In various fields, including healthcare, political science, economics,
and technology, there is a recurring challenge of identifying
important factors among many variables that may be
 associated with a
specific outcome. Such an abundance of explanatory variables often
surpasses the number of observations, making it crucial to identify
the subset of variables that genuinely affect the desired response
$Y$. This challenge is worsened by a limited understanding of how the
relevant variables influence the outcome, further emphasizing the
complexity of high-dimensional data analysis. The problem of
identifying the essential features from a pool of candidates
$\bm{X}=(X_1, \ldots , X_p)$, commonly called the variable selection
problem, can be approached by testing multiple conditional
independence hypotheses. Specifically, it involves determining which
features $X_j$ satisfy the condition $Y \not\!\perp\!\!\!\perp X_j |
\bm{X}_{-j}$, where $\bm{X}_{-j}$ represents all the measured features excluding $X_j$.

The Model-X knockoffs procedure offers a framework for selecting a set
of $X_j$ variables that are likely relevant to the response variable
$Y$, while providing finite-sample guarantees for controlling the
false discovery rate (FDR). This approach involves generating a set of
knockoff copies, denoted here as $\tilde{\bm{X}} =( \tilde{X}_1,
\ldots, \tilde{X}_p)$, which mimic the original variables and their
dependence structure, without using any information from the response
variable $Y$. This enables the differentiation between truly significant and spurious variables when comparing the original variables' importance statistics with those of their knockoff counterparts. To construct knockoff feature variables
$\tilde{\bm{X}}$,  it is crucial to accurately specify the distribution
$P_{\bm{X}}$ of the original feature variables $\bm{X}$.
Errors in the estimation or specification of
$P_{\bm{X}}$ in the knockoff procedure tend to induce an inflation of the FDR
\citep{barber2020robust}.

Various methods for determining $P_{\bm{X}}$ and generating knockoffs have been proposed and studied in the literature. The common approach utilizes a multivariate Gaussian distribution to model the joint distribution of $\bm{X}$ \citep{candes2018panning,spector2022powerful}. More flexible alternatives that are suitable for non-Gaussian scenarios include the second-order knockoff approach \citep{candes2018panning}, the Metropolized Knockoff sampler \cite{bates2021metropolized}, the Sequential knockoff algorithm for mixed data types \citep{kormaksson2021sequential}, and the Latent Gaussian Copula Knockoff procedure \citep{vasquez2023controlling}. Other relevant strategies for sampling knockoff copies adopt a likelihood-free generative approach, such as \cite{romano2020deep} and \cite{sudarshan2020deep}. \cite{kurz2022vine} proposes using simplified vine copulas to generate approximate knockoffs when dealing with a feature vector $\bm{X}$ that exhibits a complex dependence structure.

Building upon Kurz' vine foundation, we introduce the Truncated D-vine
Copula Knockoffs (TDCK) algorithm, which characterizes the joint
distribution of the vector $(\bm{X}, \tilde{\bm{X}})$ by truncating a
$2p$-dimensional D-vine copula after the $p-1$ tree level. Thus, our
proposal's novelty is using a specific vine copula structure to
approximate knockoff variables, capturing complex multivariate
dimensional structures for the covariables, and
encompassing non-linear
dependence, tail dependence, and heavy taildness. Our algorithm
significantly enhances the use of vine copulas to sample knockoffs in
three main aspects. Firstly, the truncation effectively reduces the
dependence between each original variable $X_j$ and its synthetic
counterpart $\tilde{X}_j$, yielding statistically powerful knockoffs,
particularly in high-correlation predictor scenarios. Secondly, we
propose a straightforward non-parametric approach for marginal
transformations that accommodates any continuous marginal
distribution, eliminating the need for a specific parametric family or
a kernel density estimator.

A further benefit of our approach is the use of the
\verb|rvinecopulib R| package for vine copula fitting
\citep{rvinecopulib}, which enriches the flexibility and robustness of
our method beyond what was explored in \cite{kurz2022vine}. Such a
framework includes the possibility of using an expanded range of
parametric pair-copula families (with counterclockwise density
rotations), the capability to fit non-parametric D-vine copula models,
and the ability to apply
the modified Bayesian Information Criterion
for Vines (mBICV) \citep{nagler2019model}. The mBICV is tailored for
high-dimensional scenarios and particularly helps select sparse model
classes, like truncated vine copulas (see Section
\ref{sec:Overview_Dvine}). Sparse vine models help address the problem
induced by the quadratic growth in the parameters as the dimension
increases in the vine copula.

Several runs of the Model-X knockoffs on the same dataset can result
in different sets of selected variables. This undesirable property was
addressed by \cite{ren2022derandomized}, where the
Model-X knockoffs approach was found to be equivalent to an e-value
Benjamini-Hochberg (e-BH) procedure \citep{wang2022false}. Leveraging
this connection, the authors devised
the algorithm Derandomized
Knockoffs with e-values (DKe), which mitigates the inherent randomness of
the knockoffs by aggregating the e-values obtained from multiple
runs. Accordingly,
we wrap the TDCK algorithm with the DKe method, resulting
in the proposed Derandomized Truncated D-vine Copula Knockoffs with
e-values (DTDCKe) procedure. This methodology aims to provide a stable
variable selection method under a knockoff framework
that exploits
simplified D-vine copulas to estimate the true distribution
$P_{\bm{X}}$.

In this work, we aim to understand if capturing the dependencies on the
data through a truncated D-Vine copula is enough to generate
approximate knockoffs
that adequately control the FDR. Despite the
simplifying assumption of the D-vine copulas and a specific truncated
D-vine structure to sample knockoffs, the DTDCKe procedure provides
valuable approximate knockoffs. This aspect is demonstrated in
extensive simulation studies,
with the DTDCKe procedure controling the
FDR at the prespecified level in almost every scenario. The
simulations include Gaussian and non-Gaussian feature vectors $\bm{X}$
with Gaussian and logistic linear responses. The proposed method
achieves better statistical power than the competing
methods in nearly
every simulation setup. Furthermore, to illustrate the efficacy and
relevance of the DTDCKe methodology, we apply it to a dataset
comprising gene expressions from lung tumors to identify the relevant
genes that contribute to the differentiation between normal and
cancerous tissue. This example is motivated by vine copulas being
successfully employed to construct gene networks
\cite{chatrabgoun2020constructing,farnoudkia2021vine}. Our proposed
method achieves a more reliable selection of genes than other
competing methods when the results are contrasted with an existing
meta-analysis \citep{cai2019lce}.

The paper is structured as follows. Section \ref{sec:Overview_Model-X}
presents the necessary preliminaries about the Model-X knockoffs
procedure. Section \ref{sec:Knockoffs_vine_copulas} introduces the
proposed methodology: the TDCK algorithm for sampling knockoffs using
truncated D-vines copulas. Section \ref{sec:Overview_DKe} discusses
the Derandomized Knockoffs with e-values (DKe) proposed by
\cite{ren2022derandomized}. Section \ref{sec:DTDCK} introduces the
DTDCKe procedure, which extends the TDCK methodology by wrapping it
with the DKe algorithm to mitigate the randomness of the knockoff
method's variable selection. Section \ref{sec:Simulations_paper}
presents a detailed simulation study that explores various
data-generating processes (DGPs) for the feature vector $\bm{X}$,
assuming Gaussian and logistic linear responses; here, the statistical
power and FDR are compared with those produced by Gaussian and
second-order knockoffs. Section \ref{sec:Application} illustrates the
use of the DTDCKe procedure by analyzing a gene expression dataset
from a study of lung cancer. The paper concludes with a comprehensive
discussion of the proposed methods. Appendix \ref{sec:Simulations}
provides a detailed account of two additional simulation scenarios,
which offer valuable insights into the proposed procedures' potential
efficacy.

\section{Overview of the Model-X knockoffs framework}\label{sec:Overview_Model-X}

Consider a scenario where a response variable $Y$ depends on $p$
potential explanatory covariables $\bm{X}=(X_1,X_2,\ldots,X_p )$. One
main challenge is identifying the sets of explanatory variables,
whose
joint dependence structure exhibits a genuine and meaningful
association with the response variable. This task can be viewed as a
multiple hypothesis testing procedure, wherein we aim to test the null
hypotheses $Y\indep X_j |\bm{X}_{-j}$ for each
$j\in[p]:=\{1,2,\ldots,p\}$, with $\bm{X}_{-j}=\{X_i:i\neq j\}$,
\cite{candes2018panning}. We refer to the variables with no
association with the response as null,
and to those with an association
as non-null. This study adopts the Model-X framework since it aims to
detect as many non-null variables as possible
while controlling the FDR at a pre-specified level.

Let $S_0$ be the subset of $[p]$ containing the indices of the null
variables, and $S= [p] \setminus S_0$ be the set with the indices of
the non-null or relevant variables. The FDR is defined as the
following expected value
\begin{equation}\label{Equ_FDR}
    \textrm{FDR} =\mathbb{E}\left[\frac{|\hat{S}\cap S_0|}{|\hat{S}| \vee 1}\right],
\end{equation}
where $\hat{S}$ is the set of the selected variables and $a \vee b = \textrm{max}(a,b)$. The application of Model-X knockoffs entails the construction of a series of synthetic variables $\tilde{\bm{X}} = ( \tilde{X}_1, \tilde{X}_2, \ldots, \tilde{X}_p )$, which follow two fundamental conditions:
\begin{enumerate}
    \item \textit{Exchangeability}: For any subset $R\subset [p]$, it follows that $(\bm{X},\tilde{\bm{X}})_{\textrm{swap}(R)}\stackrel{d}{=}(\bm{X},\tilde{\bm{X}})$, where $\textrm{swap}(R)$ means swapping $X_j$ and $\tilde{X}_j$ for each $j\in R$.
    \item \textit{Conditional independence}: $Y \indep \tilde{\bm{X}}|\bm{X}$.
\end{enumerate}
The exchangeability property indicates that discerning whether the $j$-th column represents a genuine variable or a knockoff should be impossible based solely on observations of $\bm{X}$ and $\tilde{\bm{X}}$. The second property asserts that knockoffs do not yield any additional information regarding the response $Y$.

Let $\bm{Y}\in \mathbb{R}^{n\times 1}$ be the vector of responses and $\bm{\mathcal{X}}\in \mathbb{R}^{n\times p}$ its corresponding design matrix of observations. The Model-X knockoff methodology requires sampling a copy $\tilde{\bm{X}} = ( \tilde{X}_1, \tilde{X}_2, \ldots, \tilde{X}_p )$ for each observation of $\bm{X}=(X_1,X_2,\ldots,X_p )$. Various methods in Gaussian and non-Gaussian settings have been proposed to sample valid knockoffs given the vector $\bm{Y}$ and the matrix $\bm{\mathcal{X}}$; we refer the reader to \cite{candes2018panning, romano2020deep, sudarshan2020deep, bates2021metropolized, kormaksson2021sequential,  kurz2022vine, spector2022powerful, vasquez2023controlling}. Let $\tilde{\bm{\mathcal{X}}}$ represent the knockoff design matrix; that is, $\tilde{\bm{\mathcal{X}}}$ is analogous to $\bm{\mathcal{X}}$ but contains the knockoff copies. After sampling the knockoff variables $\tilde{\bm{\mathcal{X}}}$, the next step involves constructing the knockoff feature statistics, represented here as $\bm{W}=(W_1, W_2,\ldots,W_p )$, by utilizing the augmented dataset $(\bm{\mathcal{X}}, \tilde{\bm{\mathcal{X}}}, \bm{Y})$; each statistic is computed as $W_j=f_j(Z_j, Z_{j+p} )$, wherein $f_j$ represents an anti-symmetric function satisfying $f_j(u,v)=-f_j(v,u)$. The quantities $Z_j$ and $Z_{j+p}$ measure the importance of the original variable $X_j$ and its copy $\tilde{X_j}$ to the response variable $Y$, respectively. The anti-symmetry of $f_j$ implies that swapping the $j$-th variable with its knockoff changes the sign of $W_j$, the so-called \emph{flip-sign property}. Notably, larger positive values of $W_j$ indicate a stronger association between $X_j$ and $Y$. Various functions for determining the importance of features with the flip-sign property have been proposed in existing literature \citep{candes2018panning, gimenez2019knockoffs}.

The last step comprises selecting the set of relevant variables $\hat{S}_{kn}$ by calculating a data-dependent threshold, denoted here by $T$. This selection is based on a specific value $\alpha_{kn} \in (0, 1)$, indicating the desired level of control of the FDR. Specifically, the choice of $\hat{S}_{kn}$ is
made through the following formulation
\begin{equation}\label{Equ_T+}
  \hat{S}_{kn} :=\left \{j: W_j \geq T \right \}, 
  \;\;\textrm{ where } \;\;T:=\textrm{min}\left \{  c>0:\frac{1+\#\{j:W_j \leq -c\}}{\#\{j:W_j \geq  c\}} \leq  \alpha_{kn} \right \}. 
\end{equation}
The process of constructing the knockoff statistic $\bm{W}$ and calculating the threshold $T$ is called the \emph{knockoff filter} \citep{barber2020robust}. \cite{candes2018panning} show that the knockoff filter controls the FDR at the desired level $\alpha_{kn}$ if there is prior knowledge of the joint distribution of the feature variables $P_{\bm{X}}$; that is, the selected set $\hat{S}_{kn}$ satisfies that $\textrm{FDR} \leq \alpha_{kn}$. In addition, \cite{barber2020robust} show that the Model-X knockoffs method is robust when there exist slight errors in estimating the true distribution $P_{\bm{X}}$. 

\section{Sampling knockoffs using truncated D-vine copulas}
\label{sec:Knockoffs_vine_copulas}

This section briefly overviews D-vine copulas and introduces the Truncated D-vine Copula Knockoffs (TDCK) procedure and its implementation in \verb|R|.

\subsection{Overview of vine copulas}
\label{sec:Overview_Dvine}

Vine copula models provide a versatile framework for characterizing intricate multivariate distributions, accommodating non-linear and asymmetrical dependencies, tail dependence, and heavy-tailedness \citep{muller2019dependence, czado2019analyzing, czado2022vine}. These copula models decompose a multivariate distribution into a product of conditional and unconditional bivariate distributions. Namely, the joint distribution of a $p$-dimensional vector $\bm{X}=(X_1, \ldots, X_p)$ is expressed as the product of $p(p-1)/2$  conditional and unconditional marginal bivariate distributions on the copula scale; as usual, the copula scale is determined by transforming the individual variables $X_j$ into uniform random variables $U_j = F_j(X_j)$, where $F_j$ represents the cumulative distribution function (CDF) for $X_j$. 

A notable advantage of the vine copula models lies in the freedom to
independently select and combine arbitrary bivariate distributions for
each component. Nevertheless, it is crucial to note that there can
exist more than one construction representing the same multivariate
distribution. \cite{bedford2002vines} introduced a graphical model
called a regular vine (R-vine) that provides a general and organized
framework to determine valid decompositions for vine copulas. In this paper, we focus on a particular case of the R-vine called D-vine
copulas.

A $p$-dimensional R-vine consists of a sequence of interconnected
trees $T_m = (V_m, E_m)$, $m = 1, \ldots, p-1$, where $V_m$ and $E_m$
denote the node and edge sets of the tree $T_m$, respectively. As
outlined by \cite{bedford2002vines}, each tree in the sequence adheres
to certain conditions. The nodes in the first tree, $T_1$, represent
the $p$ variables in the multivariate vector $\bm{X}$, while the edges
express the bivariate dependence between $p-1$ pairs of these
variables. In the second tree, $T_2$, the nodes are the edges of the
first tree, while its $p-2$ edges depict the conditional dependence
between the variables in the nodes conditional to the variable they
share; that is, $T_2$ represents the conditional dependence of $p-2$
pairs of variables. A pair of nodes in the second tree can only be
connected with an edge if there is a common variable between them, the
so-called \emph{proximity condition}. The following trees have a
similar structure, with each subsequent tree representing the
conditional dependence of a decreasing number of pairs of
variables. The last tree consists of a single edge representing the
conditional dependence of two variables, conditional on the remaining
$p-2$ variables.

To characterize the joint distribution of a vector $\bm{X} = \left(
X_1, \ldots,X_p \right)$, an R-vine copula model assigns a bivariate
conditional copula to each edge within an R-vine tree
structure. Following the description in \cite{cooke2020vine}, let
$\left [ jk|S \right ]$ be an edge in the tree $T_l$, where $S$ is a
subset of $\left \{ 1,\ldots ,p \right \}$ that does not include
$\left \{ j,k \right \}$ and has a cardinality of $l-1$. Besides, let
$F_{j}$ and $F_{j|S}$ denote the CDF of $U_j:=F(X_j)$ and $U_j$ given
$S$, respectively. An R-vine copula specifies a bivariate conditional
copula $C_{jk;S}$ to the edge $\left [ jk|S \right ]$ to couple
$F_{j|S}$ and $F_{k|S}$. When $S = \emptyset $, the copula $C_{jk;S}$
equals $C_{jk}$; i.e., it becomes the bivariate copula connecting the
CDFs $F_j$ and $F_k$ in $T_1$. The conditional dependence between two
variables may vary depending on the values taken by the conditioning
set $S$. However, to enhance modeling tractability, it is customary to
simplify an R-vine copula by assuming that the conditional dependence
is independent of $S$. Under this simplifying assumption, $C_{jk;S}$
does not depend on the values of variables indexed by $S$. Although
this supposition may not be valid in certain scenarios, simplified
R-vine copula models can still provide valuable approximations for
real-world situations when the simplifying assumption is not entirely
satisfied \citep{haff2010simplified, nagler2016evading,
  czado2022vine}.

A D-vine (drawable vine) denotes an R-vine wherein all nodes in the
initial tree $T_1$ are connected to a maximum of two other nodes. The
proximity condition implies that for the class of D-vines, the
characteristics of the tree $T_1$ govern the specifications of all
subsequent trees in the sequence. The variable order is commonly
determined by maximizing the dependence within the first tree of the
D-vine \citep{czado2019analyzing}; this is regularly done by employing
heuristic optimization procedures. This particular task can be viewed
as analogous to the traveling salesman problem, as discussed by
\cite{brechmann2010truncated}. Solving this problem requires
identifying a maximum Hamiltonian path on the weighted complete graph
of the nodes in $T_1$, with weights given by Kendall's tau association
measure between the nodes.

\subsection{Constructing knockoff variables using truncated D-vine copulas}

In this study, we extend the D-vine copula-based knockoff sampling
procedure introduced in \cite{kurz2022vine} so that the resulting
construction generates a more robust and powerful knockoff sampling
methodology. In the following, we describe in detail our proposal.

\begin{figure}[ht]
    \centering
    \includegraphics[width=1\textwidth]{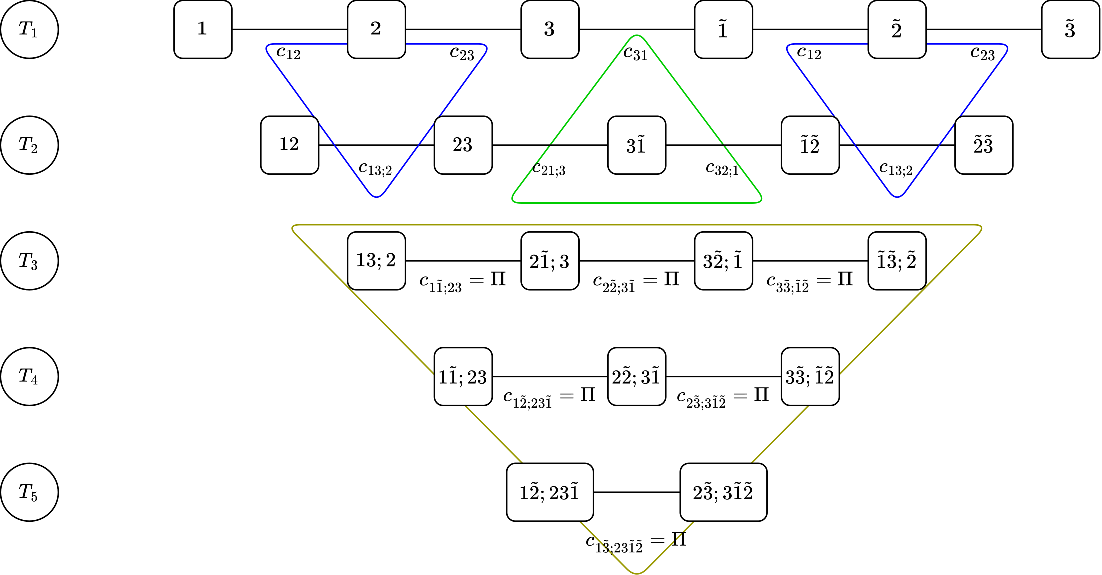}
    \caption{ Diagram of a 3-truncated 6-dimensional D-vine copula to model the vector $(X_1,X_2,X_3,\tilde{X}_1, \tilde{X}_2, \tilde{X}_3)$ for sampling knockoffs.}
    \label{Fig1}
\end{figure}

For constructing knockoff variables $\tilde{\bm{X}}$ based on D-vine copulas, the distribution of the vector $( \bm{X},\tilde{\bm{X}} )$ is represented as a $2p$-dimensional D-vine copula. We assume that the order
in $T_1$ is set to $X_1 \rightarrow X_2 \rightarrow \ldots \rightarrow
X_p \rightarrow \tilde{X}_1 \rightarrow \tilde{X}_2 \rightarrow \ldots
\rightarrow \tilde{X}_p$. To serve as an example of our construction,
consider the scenario of a 3-dimensional predictor vector $\bm{X}=
\left ( X_1, X_2, X_3\right )$. Figure \ref{Fig1} illustrates a
6-dimensional D-vine copula model for the distribution of $( X_1, X_2,
X_3, \tilde{X}_1, \tilde{X}_2, \tilde{X}_3 )$. The blue group of
bivariate conditional and unconditional copulas represents the
3-dimensional D-vine copula model for the vector $\bm{X}$. As
expected, the same dependence structure is replicated for its knockoff
counterpart $\tilde{\bm{X}}$. The green group of conditional copulas
represents the elements connecting the 3-dimensional D-vine copulas of
$\bm{X}$ and $\tilde{\bm{X}}$. The olive green group, which contains
bivariate conditional copulas from tree $T_3$ to tree $T_5$, completes
the structure.

The 6-dimensional D-vine copula model shown in Figure \ref{Fig1} can
be estimated by duplicating the data in the copula scale and employing
a standard fitting algorithm for a 6-dimensional D-vine copula. Notice
that in the first tree in the olive green group, $T_3$, the edges
represent the conditional bivariate distribution of each variable
$X_j$ and its corresponding knockoff $\tilde{X}_j$. Fitting this tree
using duplicated data results in an extreme correlation between $X_j$
and $\tilde{X}_j$, reducing the selection's power
\citep{spector2022powerful, berti2023new}. Therefore, we select the
independent copula $\Pi$ for all the conditional pair copulas in the
tree $T_3$ to make $\tilde{\bm{X}}$ as independent of $\bm{X}$ as
possible. Following this approach, the remaining conditional copulas
for trees $T_4$ and $T_5$ are also chosen as the independence copula,
as depicted in Figure \ref{Fig1}. Consequently, the configuration in
Figure \ref{Fig1} represents a 3-truncated 6-dimensional D-vine
copula, or a $p$-truncated $2p$-dimensional D-vine copula in the
general case. Truncated vine copulas are copula models that introduce
sparsity by assuming that all pair copulas are equal to the
independent copula from a specific tree level $M$.

The symmetry in the proposed $p$-truncated $2p$-dimensional D-vine
copula for modeling the joint dependence of $( \bm{X} , \tilde{\bm{X}}
)$ permits the generation of approximated knockoffs, which effectively
control the FDR at a desired level. Extensive simulation experiments
detailed in Section \ref{sec:Simulations_paper} and Appendix
\ref{sec:Simulations} provide evidence of this FDR control. Including
the $p$-truncation in the D-vine copula model reduces the association
between $\bm{X}$ and $\tilde{\bm{X}}$, thereby enhancing the method's
statistical power in scenarios where the predictors exhibit high
levels of correlation. In such instances of correlation, the
conventional Gaussian knockoff sampling procedure, which aims to
minimize the mean absolute correlation (MAC) between each $X_j$ and
$\tilde{X_j}$, significantly diminishes its statistical power
effectiveness. To address this limitation in the Gaussian framework,
\cite{spector2022powerful} propose an alternative approach that
generates more powerful knockoffs by minimizing the reconstructability
(MRC) of the original variables $\bm{X}$ through their replicas
$\tilde{\bm{X}}$. When comparing our proposed D-vine copula model with
one of the best implementations of the Gaussian MRC knockoffs, our
method consistently produces knockoffs that are either equivalent to
or superior in terms of power. These findings are detailed in our
simulation experiments of Section \ref{sec:Simulations_paper} and
Appendix \ref{sec:Simulations}.

In the implementation, we employ the \verb|R|-package \verb|rvinecopulib| to estimate the $p$-truncated $2p$-dimensional D-vine copula that models the joint dependence of $( \bm{X},\tilde{\bm{X}} )$. This package provides an interface to \verb|vinecopulib|, a \verb|C++| library designed explicitly for vine copula modeling that offers a comprehensive set of tools for statistical analyses of R-vine copula models, including parameter estimation, model selection, simulation, goodness-of-fit tests, and visualization \citep{rvinecopulib}. The \verb|rvinecopulib| package contains a wide suite of parametric bivariate copula families, including:
1) elliptical copulas, which contain the
Gaussian and $t$ distributions;
2) one-parameter Archimedean copulas,such as the Clayton, Gumbel, Frank, and Joe copulas; and 3) two-parameter Archimedean copulas, including the BB1 (Clayton-Gumbel), BB7 (Joe-Clayton),
and BB8 (Joe-Frank) copulas.
Also, the package includes counterclockwise rotations of the bivariate copula densities, expanding the range of dependencies within the previous parametric families; namely, the following transformations can be utilized: $c_{90}(u_1, u_2):= c(1 - u_2, u_1)$, $c_{180}(u_1, u_2) := c(1 - u_1, 1 - u2)$, and $c_{270}(u_1, u_2) := c(u_2, 1 - u_1)$ \citep{czado2019analyzing}.

The \verb|rvinecopulib| package implements the modified Bayesian information criterion for vines (mBICV), a criterion specifically designed to select sparse vine copula models \citep{nagler2019model}. These sparse models offer computational advantages since they reduce the number of bivariate copulas to estimate. Using the mBICV mitigates the risk of overfitting and reduces the computational burden compared to classical criteria such as the AIC or the BIC when selecting pair-copula families. These features make the mBICV particularly suitable for high-dimensional scenarios. Lastly, the \verb|rvinecopulib| package allows fitting non-parametric bivariate copulas using transformed kernel estimators \citep{nagler2016evading}. The non-parametric estimation of bivariate copulas is achieved through standard kernel estimators by transforming the data from the copula scale $U_j = F_j(X_j)$ to the standard normal scale $Z_j = \Phi^{-1}(F_j(X_j))$, where $\Phi$ is the cumulative distribution function (CDF) of a standard normal distribution. This approach offers advantages in high-dimensional settings since, unlike other classical non-parametric estimators, its convergence rate is not affected by the dimensionality of the data \citep{nagler2016evading}.

There are some differences between our proposal and the one described in \cite{kurz2022vine}. The first difference is the use of truncated D-vine copulas; as we show below, with extensive simulations, this aspect is crucial for improving the statistical power in heavy-tailed scenarios. In addition, our proposal is based on a non-parametric approach for marginal transformations, where the knockoffs are sampled on the copula scale. This allows for accommodating any continuous distribution, thus making parametric models optional. The last significant difference is that we exploit the rich options of the \verb|R| package \verb|rvinecopulib|, which allows the use of parametric and non-parametric copulas and many other alternatives.

\subsection{Truncated D-vine Copula Knockoffs Algorithm}

Let $\bm{\mathcal{X}} = (\bm{x}_1^T, \ldots , \bm{x}_p^T) \in \mathbb{R}^{n \times p}$ be a matrix with $n$ observed data points, where $\bm{x}_j^T$ represents a column vector with the $n$ observations of the variable $X_j$. Similarly, let $\tilde{\bm{\mathcal{X}}}=  (\tilde{\bm{x}}_{1}^T, \ldots, \tilde{\bm{x}}_{p}^T)\in \mathbb{R}^{n \times p}$ denote the matrix of sampled knockoff points, with $\tilde{\bm{x}}_j^T$ representing a column vector with the $n$ realizations of the variable $\tilde{X}_j$. Our proposal to sample knockoffs $\tilde{\bm{\mathcal{X}}}$ involves three main steps. Firstly, a $p$-truncated $2p$-dimensional D-vine copula is fitted to represent the joint dependency structure of $( \bm{X},\tilde{\bm{X}} )$. Secondly, the knockoff variables are sampled on the copula scale employing the estimated D-vine copula. We denote these knockoff variables by $\tilde{\bm{\mathcal{U}}} = (\tilde{\bm{u}}_{1}^T, \ldots, \tilde{\bm{u}}_{p}^T) \in \mathbb{R}^{n \times p}$; here, $\tilde{\bm{u}}_j^T$ represents a column vector with the $n$ observations of the variable $\tilde{U}_j=F_j(\tilde{X_j}) \sim \textrm{Unif}[0,1]$. The last step involves pulling the sampled knockoff variables $\tilde{\bm{\mathcal{U}}}$ to their original scale $\tilde{\bm{\mathcal{X}}}$. In the following, we describe the central aspects associated with these three steps.

The first stage to estimate the $p$-truncated $2p$-dimensional D-vine copula model for the vector $(\bm{X}, \tilde{\bm{X}})$ consists of selecting the order of the variables in the first tree, which can be achieved by employing the \verb|TSP| \verb|R| package \citep{TSP}. Once the $p$-dimensional D-vine tree structure for $\bm{X}$ is chosen, the features matrix $\bm{\mathcal{X}}$ is transformed to the copula scale $\bm{\mathcal{U}} = (\bm{u}_1^T, \ldots, \bm{u}_p^T)$, $\bm{u}_j^T=(F_j(x_{1j}), \ldots, F_j (x_{nj}))^T$; here, the column vector $\bm{u}_j^T$ represents the $n$ realizations of the random variable $U_j=F_j(X_j)\sim \textrm{Unif}[0,1]$. This transformation is performed by computing the pseudo-observations associated with $\bm{X}$. Given a sample $(t_1, \ldots, t_m)$ of a random variable $T$, the pseudo-observations of $T$ are given by $u_i = r_i / (m+1)$, where $r_i$ is the $i$-th entry of the sample's rank statistic $(r_1,r_2,\ldots,r_m)$. It follows that the pseudo-observations can be computed by applying the empirical distribution function of $T$ to the data and then scaling the resulting values by $m/(m+1)$. The scaling factor, which is asymptotically negligible, is used to confine the variables within the interval $[0,1]$ to avoid issues at the boundaries. The transformation of a sample to pseudo-observations is applied using the \verb|rvinecopulib| package through the function \verb|psedo_obs()|. We employ this function to convert the features matrix $\bm{\mathcal{X}}$ to a pseudo-observations matrix.

The pseudo-observations are duplicated to generate a $2p$-dimensional dataset $\bm{\mathcal{U’}}$ through $\bm{\mathcal{U’}} = (\bm{u}_1^T, \ldots, \bm{u}_p^T, \bm{u}_1^T, \ldots, \bm{u}_p^T) \in \mathbb{R}^{n \times 2p}$. Next, a $p$-truncated $2p$-dimensional D-vine copula is fitted to $\bm{\mathcal{U’}}$. The function \verb|vinecop()| from the \verb|rvinecopulib| package can be used for this task,
with some of its arguments accepting helpful options for estimating the $p$-truncated $2p$-dimensional D-vine copula. The argument \verb|family_set| allows for the specification of the options \verb|parametric| or \verb|non-parametric|, depending on the desired approach. If the argument \verb|par_method| is set to \verb|mle|, the copula is fitted using maximum likelihood estimation. The \verb|selcrit| argument indicates the criterion employed for model selection, and its \verb|mbicv| option ensures sparsity in high dimensions, as discussed above. Lastly, the argument \verb|trunc_lvl| should be set to the desired truncation level $p-1$ to achieve truncation after tree $T_{p-1}$. This concludes the description of the first step in our knockoff procedure.

The following step is to sample knockoff variables $\tilde{\bm{\mathcal{U}}}$ in the copula scale; for this task, we utilize the Rosenblatt transform $R(\cdot)$  \citep{czado2022vine}. The Rosenblatt transform $R(\cdot)$ converts a random vector $\bm{U} = (U_1, \ldots , U_p)$ following a D-vine copula into a vector $\bm{V} = (V_1, \ldots, V_p) = R(\bm{U})$ consisting of independent $\textrm{Unif}[0,1]$ random variables. Conversely, the inverse operation $\bm{U} = R^{-1}(\bm{V})$ transforms independent uniform random variables $\bm{V}$ into a vector $\bm{U}$ characterized by the D-vine copula associated with $R(\cdot)$. We propose the iterative process described in the following paragraph to generate knockoffs on the copula scale $\tilde{\bm{\mathcal{U}}}$ using these transformations.

The Rosenblatt transformation is applied to the first row $(u_{11}, \ldots, u_{1p})$ in the transformed data matrix $\bm{\mathcal{U}} = (\bm{u}_1^T, \ldots, \bm{u}_p^T)$, $\bm{u}_i^T=(F_i(x_{1i}), \ldots, F_i (x_{ni}))^T$. This requires the $p$-dimensional D-vine copula model for the vector $\bm{X}$,
as illustrated in blue in Figure \ref{Fig1}. Extracting this model from the whole $p$-truncated $2p$-dimensional D-vine copula is straightforward. Applying the Rosenblatt transform to $(u_{11}, \ldots, u_{1p})$ generates a new vector $(v_{11}, \ldots, v_{1p})$, representing a realization of vector $\bm{V}$. Subsequently, $p$ independent realizations of a uniform $\textrm{Unif}[0,1]$ random variable are sampled and arranged in a vector $(\tilde{v}_{11}, \ldots, \tilde{v}_{1p})$; then, the inverse Rosenblatt transform for the estimated $p$-truncated $2p$-dimensional D-vine copula estimated from $\bm{\mathcal{U}}'$ is applied to vector $(v_{11}, \ldots, v_{1p}, \tilde{v}_{11}, \ldots, \tilde{v}_{1p})$, which yields a realization of a $2p$-dimensional vector $(u_{11}, \ldots, u_{1p}, \tilde{u}_{11}, \ldots, \tilde{u}_{1p})$ that follows the $p$-truncated $2p$-dimensional D-vine copula. The last $p$ entries of this $2p$-dimensional vector constitute the knockoff sample in the copula scale for the first observation. Repeating this procedure for the remaining $n-1$ rows in $\bm{\mathcal{U}}$ produces the knockoff matrix $\tilde{\bm{\mathcal{U}}}$. The Rosenblatt transform for a truncated D-vine copula can be computed using the package \verb|rvinecopulib| via the function \verb|rosenblatt()|, while its inverse can be obtained using the function \verb|inverse_rosenblatt()|.

The final step involves pulling the matrix $\tilde{\bm{\mathcal{U}}}$ into the scale of the original marginal distributions, where each $\tilde{\bm{u}}_j^T$ is transformed using the quantile function $F_j^-$ of the original variable $X_j$. Consequently, it is necessary to compute an estimator of the quantile function $F_j^-$. In this study, we adopted the quantile estimators $\hat{Q}^{(k)}(s)$, $k=1, 2, \ldots,9$, as described in \cite{hyndman1996sample}; specifically, we select the estimator $\hat{Q}^{(k)}(s)$ with $k=8$ since it provides an approximately median-unbiased estimator, which can be defined independently of the underlying distribution, and possesses most of the desirable properties of a quantile estimator. For each $j$, the knockoff sample in the original scale $\tilde{\bm{x}}_{j}^T$ is obtained by applying the sample quantile function $\hat{Q}^{(8)}(s)$ of $X_j$ to the vector $\tilde{\bm{u}}_j^T$. The \verb|quantile()| function in the \verb|stats| \verb|R| package is used for the computation of these sample quantile functions.

Algorithm 1 summarizes the steps and substeps required to sample approximated knockoffs employing the $p$-truncated $2p$-dimensional D-vine copula that models the joint distribution of $(\bm{X}, \tilde{\bm{X}})$. We refer to such an algorithm as the Truncated D-vine Copula Knockoffs (TDCK) algorithm.

\par\noindent\rule{\textwidth}{1pt}

\textbf{Algorithm 1}: Truncated D-vine Copula Knockoffs
\par\noindent\rule{\textwidth}{0.5pt}
\textbf{Input}: A set of $n$ independent samples $(X_{i1}, \ldots , X_{ip})$ arranged in a data matrix $\bm{\mathcal{X}} \in \mathbb{R}^{n \times p}$.

\begin{enumerate}

    \item Fit a $p$-truncated $2p$-dimensional D-vine copula:
    
    \begin{enumerate}
        \item Determine the order of the variables in the first tree $T_1$ for a $p$-dimensional D-vine tree structure.
        \item Transform the columns of $\bm{\mathcal{X}}$ to the copula scale $\bm{\mathcal{U}}\in \mathbb{R}^{n \times p}$ using the pseudo-observations approach.
        \item Duplicate the data on the copula scale $\bm{\mathcal{U}}$ to create a $2p$-dimensional dataset $\bm{\mathcal{U’}}\in \mathbb{R}^{n \times 2p}$.
        \item Fit a $p$-truncated $2p$-dimensional D-vine copula model to $\bm{\mathcal{U’}}$ using the mBICV pair-copula selection criterion and maximum likelihood estimation.
       
    \end{enumerate}

   \item Sample the matrix of knockoffs in the copula scale $\tilde{\bm{\mathcal{U}}} \in \mathbb{R}^{n \times p}$ by performing the following steps for each $i = 1, 2, \ldots, n$:
    \begin{enumerate}
        \item Apply to the $i$-th row of $\bm{\mathcal{U}}$ the Rosenblatt transform corresponding to the fitted $p$-dimensional D-vine copula model for $\bm{{X}}$, which generates the  vector  $({v}_{i1}, \ldots, {v}_{ip})$ of i.i.d. $\textrm{Unif}[0,1]$ random variables.
        \item Simulate a $p$-dimensional random vector $(\tilde{v}_{i1}, \ldots, \tilde{v}_{ip})$ of i.i.d. $\textrm{Unif}[0,1]$ random variables.
        \item Apply the inverse of the Rosenblatt transform of the previously fitted $p$-truncated $2p$-dimensional D-vine copula to the concatenated vector $(v_{i1}, \ldots, v_{ip}, \tilde{v}_{i1}, \ldots, \tilde{v}_{ip})$, which results in a $2p$-dimensional vector $(u_{i1}, \ldots, u_{ip}, \tilde{u}_{i1}, \ldots, \tilde{u}_{ip})$; here, the last $p$ entries correspond to the knockoff sample on the copula scale for the $i$-th observation.
        
    \end{enumerate}
    
    \item Apply the sample quantiles $\hat{Q}^{(8)}(s)$ for each $X_j$ to the corresponding columns of $\tilde{\bm{\mathcal{U}}}$, yielding the knockoff samples in their original scale.
\end{enumerate}

\noindent\textbf{Output}: A set of $n$ independent knockoff samples $(\tilde{X}_{i1}, \ldots , \tilde{X}_{ip})$ arranged in a data matrix $\tilde{\bm{\mathcal{X}}}\in \mathbb{R}^{n \times p}$.
\par\noindent\rule{\textwidth}{0.5pt}

\section{Derandomizing knockoffs using e-values}\label{sec:Overview_DKe}

According to the conceptualization in \citep{candes2018panning}, the Model-X framework requires generating realizations of the synthetic variables $\tilde{\bm{X}}$ in order to apply the corresponding knockoff filter. Due to the random nature of drawing the variables $\tilde{X}_j$'s, the resulting set of selected variables $\hat{S}_{kn}$ is also subject to randomness. In other words, running the Model-X framework multiple times on the same dataset can lead to different selected sets $\hat{S}_{kn}$. Empirical evidence suggests that the output may exhibit significant variability across different runs, which could be considered an undesirable characteristic.

\cite{ren2022derandomized} proposed an algorithm called Derandomized Knockoffs with e-values (DKe) to address this issue. The e-values constitute a recent valuable approach for statistical inference and multiple hypothesis testing, closely related to betting, likelihood ratios, and Bayes factors \citep{ vovk2021values}. In the context of statistical hypothesis testing, an e-value refers to a non-negative random variable $E$ such that $\mathbb{E}[E] < 1$ when the null hypothesis is true. In analogy to p-values, the null hypothesis is rejected when the e-value surpasses a specified threshold. For instance, if $\alpha$ is the desired significance level, the null hypothesis can be rejected when $ E\geq 1/\alpha$, ensuring that the probability of committing a type-I error is less than or equal to $\alpha$ \citep{ren2022derandomized}.

Consider the null hypotheses $H_j$, $j =1,2,\ldots,p$. Let $e_j$ be the e-value from testing the null hypothesis $H_j$, $j =1,2,\ldots,p$. \cite{wang2022false} propose the e-BH procedure, which extends the Benjamini-Hochberg (BH) algorithm for controlling the FDR \citep{benjamini1995controlling}. The e-BH procedure employs the e-values $e_1, e_2, \ldots, e_p$ to control the FDR at a pre-specified level $\alpha_{ebh}$ in the multiple-hypothesis testing of $H_1,\ldots,H_p$. Analogously to the BH procedure, the e-BH procedure determines the selected set of discoveries $\hat{S}_{ebh}$ according to
\begin{equation}\label{Equ_e-bh}
    \hat{S}_{ebh}=\left \{ j: e_j \geq \frac{p}{\alpha \hat{k}} \right \}, \;\;  \;\; \hat{k} = \textrm{max}\left \{ k\in \left [ p \right ] : e_{(k)}\geq \frac{p}{\alpha k}  \right \},
\end{equation}
where $e_{(j)}$ denotes the $j$-th largest value among $e_1, e_2, \ldots, e_p$ and $\left [ p\right ]:=\{1,2,\ldots,p\}$. \cite{wang2022false} prove that the e-BH procedure controls the FDR by showing that  FDR$\leq \alpha |S_0|/p \leq \alpha$ for any configuration of e-values $e_j$, regardless of their statistical dependence.

The key principle that enables employing the e-BH procedure to mitigate the effects of randomness in the Model-X knockoffs lies in the connection between these two approaches. \cite{ren2022derandomized} prove that the Model-X knockoffs can be reinterpreted as an e-BH procedure; specifically, $\hat{S}_{kn}$, the set of selected variables by the knockoff filter, is equivalent to the set $\hat{S}_{kn-ebh}$ of variables selected by the e-BH procedure when the e-values considered satisfy
\begin{equation}\label{Equ_e-j}
    e_j := p\cdot \frac{ \mathbb{I} \left \{ W_j\geq T \right \} }{1+\sum_{k \in [p]} \mathbb{I}\left \{ W_k \leq  -T \right \} },
\end{equation}
where $W_i$ is the knockoff feature statistic and $T$ is the data-dependent threshold of the knockoff filter (Section \ref{sec:Overview_Model-X}).

The DKe algorithm is a procedure that requires $M$ iterations. In each iteration $m \in \left \{1, 2,\ldots, M\right \}$, a knockoff filter with a target level $\alpha_{kn} \in (0,1)$ is executed, and the e-values $e_1^{(m)},\ldots,e_p^{(m)}$ for each run $m$ are computed using equation \ref{Equ_e-j}. After completing the iterations, the average e-value for each variable $j \in [p]$ is calculated as $e_j^{avg}= \frac{1}{M}\sum_{m=1}^{M}e_j^{(m)}$. Subsequently, the e-BH procedure is applied to the averaged e-values $e_1^{avg}, \ldots , e_p^{avg}$ using a distinct target level $\alpha_{ebh} \in (0,1)$. This last step completes the derandomized knockoff procedure and ensures control of the FDR at level $\alpha_{ebh}$. It is worth mentioning that while parameter $\alpha_{ebh}$ establishes the ultimate FDR guarantee of the derandomized knockoff procedure, parameter $\alpha_{kn}$ significantly influences the method's statistical power. Empirical findings by \cite{ ren2022derandomized} suggest that selecting $\alpha_{kn}=\alpha_{ebh}/2$ leads to suitable power results. Nevertheless, determining the optimal value for $\alpha_{kn}$ remains an open question.

\section{ Derandomized Truncated D-vine Copula Knockoffs with e-values procedure}\label{sec:DTDCK}

As mentioned before, we wrap the TDCK algorithm with the DKe method to produce the Derandomized Truncated D-vine Copula Knockoffs with e-values (DTDCKe) procedure. This methodology consists of employing the TDCK algorithm in each iteration of the DKe algorithm. The resulting method mitigates the randomness of the variable selection performed by the knockoff filter based on the TDCK algorithm. The step-by-step process of the DTDCKe procedure is outlined in Algorithm 2.

\par\noindent\rule{\textwidth}{1pt}

\textbf{Algorithm 2}: Derandomized Truncated D-vine Copula Knockoffs with e-values (DTDCKe) procedure
\par\noindent\rule{\textwidth}{0.5pt}
\textbf{Input}: A set of $n$ independent samples $(X_{i1}, \ldots , X_{ip})$ arranged in a data matrix $\bm{\mathcal{X}}\in \mathbb{R}^{n \times p}$; parameters $\alpha_{kn} \in (0,1) $ and $\alpha_{ebh} \in (0,1)$.

\begin{enumerate}
    \item For each $m= 1, \ldots, M$:
    \begin{enumerate}
        \item Sample knockoffs $\tilde{\bm{\mathcal{X}}}^{(m)}$ using the TDCK algorithm (\textbf{Algorithm 1}). 
        \item Compute the vector of knockoff feature statistics $\bm{W}^{(m)}$.
        \item Calculate the knockoff threshold $T^{(m)}$ for a control level $\alpha_{kn}$ using Eq. (\ref{Equ_T+}) from Section \ref{sec:Overview_Model-X}.
        \item Compute the e-values $e_1^{(m)},\ldots,e_p^{(m)}$ applying Eq. (\ref{Equ_e-j}) from Section \ref{sec:Overview_DKe}. 
    \end{enumerate}

    \item Apply the e-BH procedure to the averaged e-values for a control level $\alpha_{ebh}$ using Eq. (\ref{Equ_e-bh}) from Section \ref{sec:Overview_DKe}.
\end{enumerate}

\noindent\textbf{Output:} The selected set of variables $\hat{S}_{kn-ebh}$ that control the FDR at $\alpha_{ebh}$ .
\par\noindent\rule{\textwidth}{0.5pt}

The DTDCKe algorithm has two forms: the parametric DTDCKe algorithm and the non-parametric DTDCKe algorithm, depending on how the truncated D-vine copula is estimated. Although the methodology's core is the same for both approaches, we refer to them as different methods to compare the benefit of each fitting approach.

The DTDCKe algorithm must be combined with an importance statistic $\bm{W}$ (see Section \ref{sec:Overview_Model-X}) to have all the elements required to employ the Model-X framework. In this research, we propose assessing the importance $Z_j$ of each variable utilizing the absolute values of the estimated coefficients obtained from a regularized regression model based on Lasso (Least Absolute Shrinkage and Selection Operator) \citep{tibshirani1996regression}. Thus, we use the Lasso coefficient-difference (LCD) statistic as knockoff feature statistics \citep{candes2018panning}, which is defined by $W_j=Z_j-Z_{j+p}=|\beta_j|-|\beta_{j+p}|$.

The regularization parameter $\lambda$ necessary to estimate the coefficients for the LCD statistic is computed through cross-validation (CV) \citep{goeman2010l1}. We adopt the percentile-Lasso method \citep{roberts2014stabilizing} to address the sensitivity of fold assignment and prevent instability in model selection. This methodology involves fitting the penalized regression model $M_{Lasso}$ times. The resulting different values for the tuning parameter $\lambda$, denoted here as $\bm{\Lambda}(M_{Lasso}) =(\hat{\lambda}_1,...,\hat{\lambda}_{M_{Lasso}})$, are then utilized to determine the $\theta$-percentile of $\bm{\Lambda}(M_{Lasso})$. Notably, the median $\hat{\lambda}(50)$ corresponds to the standard Lasso. As suggested by \cite{roberts2014stabilizing}, we employ $M_{Lasso}$ values starting at $M_{Lasso}=10$ to ensure a satisfactory performance.

\section{Simulation study}
\label{sec:Simulations_paper}

In this section, we conduct simulations to assess the DTDCKe procedure's finite sample performance. Specifically, we compare the statistical power and FDR control of five distinct Model-X knockoffs: the parametric DTDCKe procedure, the non-parametric DTDCKe procedure, Vineknockoffs \citep{kurz2022vine}, Gaussian knockoffs, and second-order knockoffs. For the Gaussian knockoffs, we employ the minimum variance-based reconstructability (MVR) knockoffs  described in \cite{spector2022powerful}. The MVR Gaussian knockoffs exhibit greater power in high correlation scenarios than the standard Gaussian knockoffs. The second-order knockoffs constitute a viable approach for cases where the vector $\bm{X}$ follows a non-Gaussian distribution. Instead of constraining $(\bm{X},\tilde{\bm{X}})_{\textrm{swap}(R)}\stackrel{d}{=}(\bm{X},\tilde{\bm{X}})$ for any subset $R\subset \{1, \ldots, p\}$, this procedure only requires that $(\bm{X},\tilde{\bm{X}})_{\textrm{swap}(R)}$ and $(\bm{X},\tilde{\bm{X}})$ possess identical first two moments, implying the same mean and covariance \citep{candes2018panning}.

In this work, we investigate four distinct data-generating processes (DGPs) for the predictors $\bm{X}$. These include a t-tailed Markov chain, a truncated D-vine copula, a multivariate normal distribution, and a parametric D-vine copula fitted to gene expression data. The results for the last two DGPs are included in Appendix \ref{sec:Simulations}, where specific aspects of their simulations are also discussed. 

Each DGP is considered under both Gaussian and logistic linear response models. The number of non-null features equals 20\% of the $p$ covariates, with the corresponding indices randomly sampled for each simulated dataset. The non-null coefficients are drawn from a uniform distribution $\textrm{Unif}\left(\left[ -\delta, -\frac{\delta}{2}\right] \cup \left[ \frac{\delta}{2}, \delta\right] \right)$ to emulate various positive and negative effects. Following \cite{ren2022derandomized} and \cite{romano2020deep}, we set $\delta =  \alpha / \sqrt{n} $, where the amplitude $\alpha$ takes different values depending on the DGP and the regression response.

We simulate $n_{sim} = 100$ datasets for each regression model in each DGP. To ensure that Vineknockoffs, MVR Gaussian, and the second-order filters are comparable with the DTDCKe procedures, we derandomized them using the DKe method. Our target FDR level is set to 20\%; thus, we set $\alpha_{kn}=0.1$ and $\alpha_{ebh}=0.2$. We employ 5-fold cross-validation to optimize the regularization parameter $\lambda$ and fix $M_{lasso}=10$; in this manner, we try to balance computational efficiency and result robustness. We sample $M=50$ knockoffs when applying the DKe algorithm. The following subsection provides further details regarding specific aspects of the simulation experiments for the DGP considered.

\subsection{Heavy-tailed Markov chain}

The DGP considered in this subsection is non-Gaussian. Specifically, we examine the t-tailed Markov chain extensively discussed in \cite{bates2021metropolized} and in \cite{spector2022powerful}. For $R_j \sim t_\nu$ i.i.d. $j=1,\ldots,p$ and $X_1 = \sqrt{\frac{\nu-2}{\nu}}R_1$, the $(j+1)$-value of the t-tailed Markov chain is given by $X_{j+1}=\rho_jX_j + \sqrt{1-\rho_j^2}\sqrt{\frac{\nu-2}{\nu}}R_{j+1}$. We maintain constant values of $\rho_j\equiv \rho$ for the consecutive variables and take $n=300$ samples and $p=100$ features. The amplitude coefficient is set to $\alpha=8$ to ensure an appropriate signal strength. We set the correlation coefficient $\rho$ to 0.8 to create a highly correlated environment. Moreover, we introduce variations in the degrees of freedom $\nu$ to observe how changes in heavy-tailedness affect the performance of the knockoff filters. This DGP can be easily generated using the \verb|sample_ar1t()| function from the \verb|Python| package \verb|Knockpy| \citep{spector2022powerful}.

\begin{figure}[ht]
    \centering
    \includegraphics[width=1\textwidth]{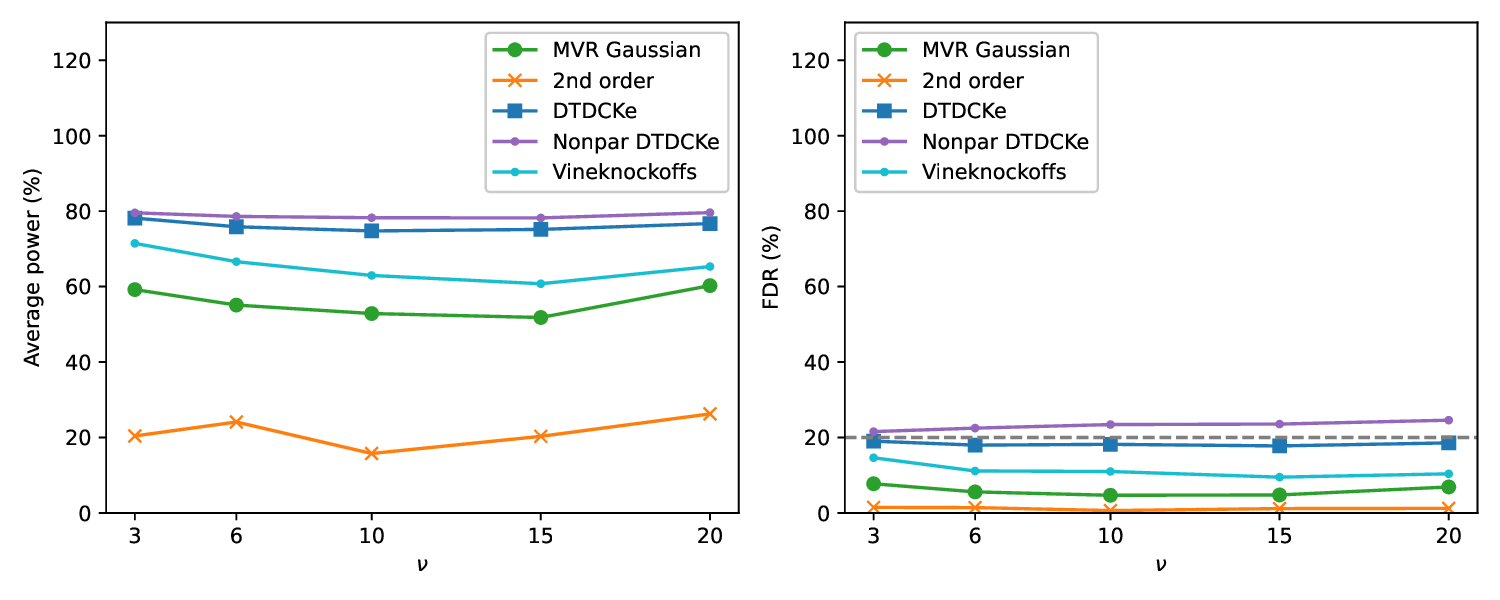}
    \caption{Empirical power and FDR as $\nu$ varies in the t-tailed Markov chain DGP for a Gaussian linear response. Each point in the graphs represents the average value across 100 repetitions.}
    \label{Fig3}
\end{figure}

Figure \ref{Fig3} presents the findings of the simulations for the Gaussian response for the DGP under consideration. Both DTDCKe procedures, parametric and non-parametric, exhibit a good statistical power performance, achieving an approximate 80\% power across all degrees of freedom $\nu$. Notably, the non-parametric approach exhibits a slightly superior performance compared to its parametric counterpart. In contrast, the Vineknockoffs method shows power metrics hovering around 70\% and 60\%, notably lower than those achieved by the DTDCKe procedures. Furthermore, the MVR Gaussian and second-order knockoff filters exhibit inferior statistical power, with values approximately at 60\% and 20\%, respectively. Remarkably, all methods effectively control the FDR at the predetermined level, although the non-parametric DTDCKe procedure shows slight inflation. There is no apparent decrease in performance, neither in power nor in FDR control, across all methods as the t-tailed Markov chain increases its heavy-tailedness.

\begin{figure}[ht]
    \centering
    \includegraphics[width=1\textwidth]{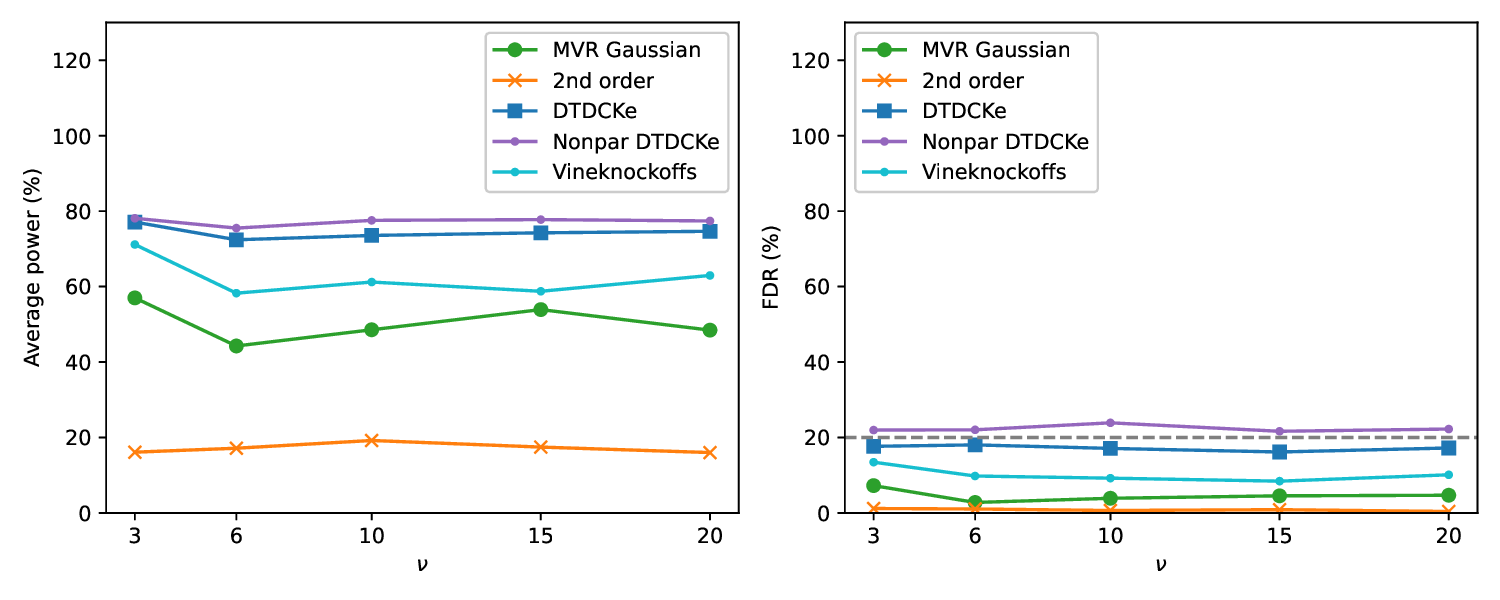}
    \caption{Empirical power and FDR as $\nu$ varies in the t-tailed Markov chain DGP for a logistic linear response. Each point in the graphs represents the average value across 100 repetitions.}
    \label{Fig7}
\end{figure}

The configuration setups for the logistic response are the same as for the Gaussian response, except for the amplitude parameter $\alpha$, which is now adjusted to $\alpha=50$. Figure \ref{Fig7} displays the results for the logistic linear response. The patterns exhibited are analogous to those observed for the Gaussian linear regression in Figure \ref{Fig3}, indicating that both DTDCKe procedures have the highest power metrics among the considered methods, with slight FDR inflation for the nonparametric case.

\subsection{Truncated D-vine copulas}\label{sec:TDcopula_sim}

For the DGP scenario where covariables $\bm{X}$ involve truncated D-vine copulas, all the pair copulas are selected from
the same family before truncation. In this study, we explore four one-parameter bivariate copula families, namely Clayton,
Frank, Gumbel, and Joe, and three two-parameter bivariate copulas,
namely t, BB1, and BB7. Kendall's $\tau$ is set to $\tau=0.7^i$ for
all the bivariate copulas in the $i$-th tree. This exponential decay
pattern of dependence is truncated when $\tau<0.1$. Consequently,
after the sixth tree, all bivariate conditional copulas become
independent bivariate copulas $\Pi$. The corresponding bivariate
copula parameters are adjusted to align with each tree's specified
value of $\tau$, thereby establishing the desired dependence
structure. We select such parameters based on a heuristic procedure
for cases where there is no bijective relationship between $\tau$ and
the copula parameters. For this DGP, we take $n=300$ samples and
$p=100$ covariates.

In this particular DGP scenario,
we incorporate skewed and heavy-tailed continuous marginal
distributions. Here, we adopt the
skew-t distribution characterized by the four-parameter family of
densities $f(x;\mu,\omega^2,\alpha, \nu)$, $x\in\mathbb{R}$
\citep{kim2003moments}, where $\mu \in \mathbb{R}$ is the location
parameter, $\omega^2 >0$ is the scale parameter, $\alpha \in
\mathbb{R}$ is a shape parameter, and $\nu>0$ denotes the degrees of
freedom. We set $\mu=0$, $\omega^2=1$, $\alpha=4$, and $\nu=5$ for our
analysis. This configuration of parameters results in a skewness value
of $\gamma_1=2.29$ and a kurtosis of $\gamma_2=15.46$, yielding
moderately skewed heavy-tailed marginal distributions.

\begin{figure}[ht]
    \centering
    \includegraphics[width=1\textwidth]{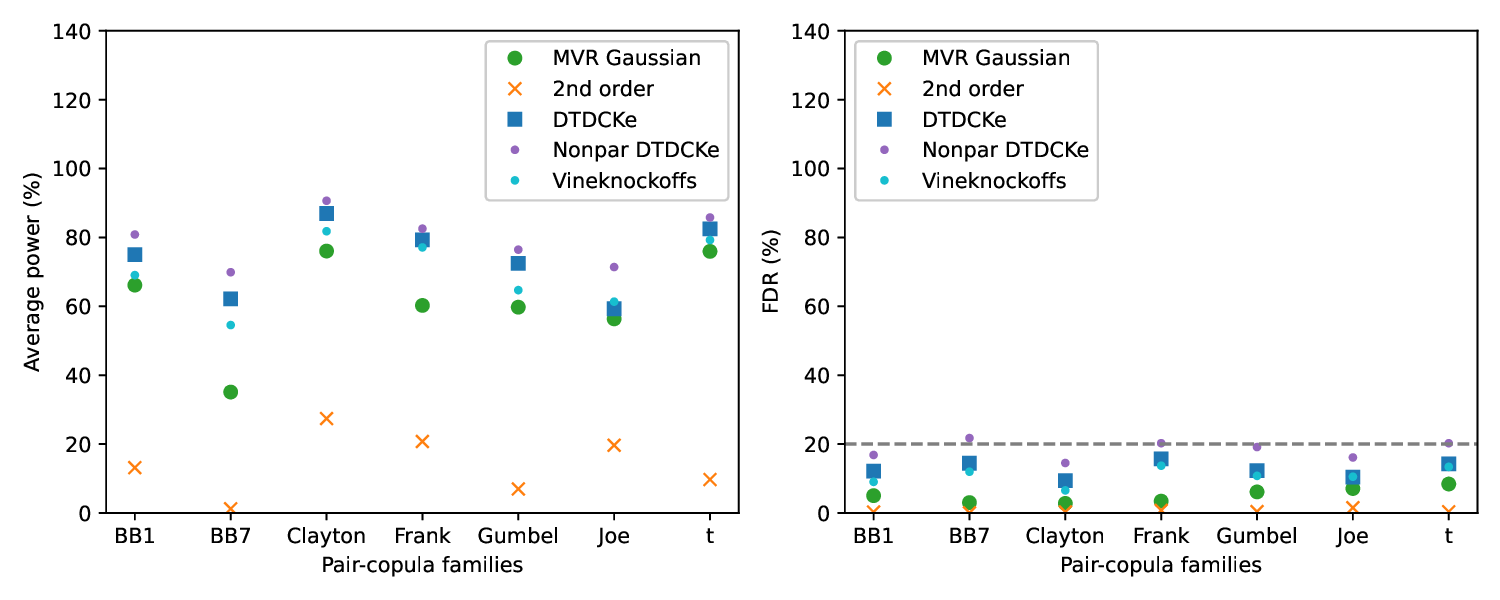}
    \caption{Empirical power and FDR of various copula families within the Truncated D-vine copulas DGP for a Gaussian linear response. Each data point in the graphs represents the average value obtained from 100 repetitions.}
    \label{Fig4}
\end{figure}

Figure \ref{Fig4} illustrates the power and FDR for the Gaussian
linear response across different bivariate copula families. We varied
the amplitude $\alpha$ within each copula family to ensure a reliable
signal, assigning values as follows: $\alpha=15$ for the Clayton and
Frank copulas, $\alpha=25$ for the Gumbel, t, and Joe copulas, and
$\alpha=20$ for the BB1 and BB7 copulas.  Both DTDCKe procedures
exhibit satisfactory statistical power, with the nonparametric method
demonstrating the highest power across all copula families. The
Vineknockoffs method also shows good power performance; however, its
power metrics are generally lower than those of the DTDCKe procedures
except for the Joe copula, in which it slightly surpasses the
parametric DTDCKe procedure. The MVR Gaussian knockoff filter
demonstrates lower power than the previous methods, except in the Joe
copula family, where its performance is nearly equivalent to those of
the parametric DTDCKe procedure and the Vineknockoffs. The
second-order knockoff filter consistently displays poor power
performance, with values falling below 30\%. Notably, all four methods
effectively control the FDR at the predetermined level for each copula
family considered. However, the non-parametric DTDCKe procedure shows
a slight inflation of the FDR for the BB7 bivariate copula family.

For the logistic response, some configuration setups differ from those of the Gaussian response case. We  decided to increase the amplitude to $\alpha=100$ to ensure sufficient statistical power. Besides, the non-null coefficients no longer follow a uniform distribution $\textrm{Unif}\left(\left[ -\delta, -\frac{\delta}{2}\right] \cup \left[ \frac{\delta}{2}, \delta\right] \right)$ as for the Gaussian response. Employing skewed and heavy-tailed continuous marginal distributions can lead to class imbalance issues in the simulated datasets. Specifically, the proportion of instances labeled as `1' may fall below 20\% or surpass 80\%. Thus, we modify the configuration to maintain a proportion of 1s close to 50\% as follows: 1) half of the non-null coefficients are now sampled from a uniform distribution $\textrm{Unif}\left( \left[ \frac{\delta}{2}, \delta\right] \right)$; 2) the remaining half of the non-null coefficients are identical to the previous coefficients but with a negative sign. As before, the vector location of these non-null coefficients is randomly sampled in each simulated dataset. This adjustment achieves a class balance close to 50\%. The rest of the configuration remains the same as for the Gaussian response.

\begin{figure}[ht]
    \centering
    \includegraphics[width=1\textwidth]{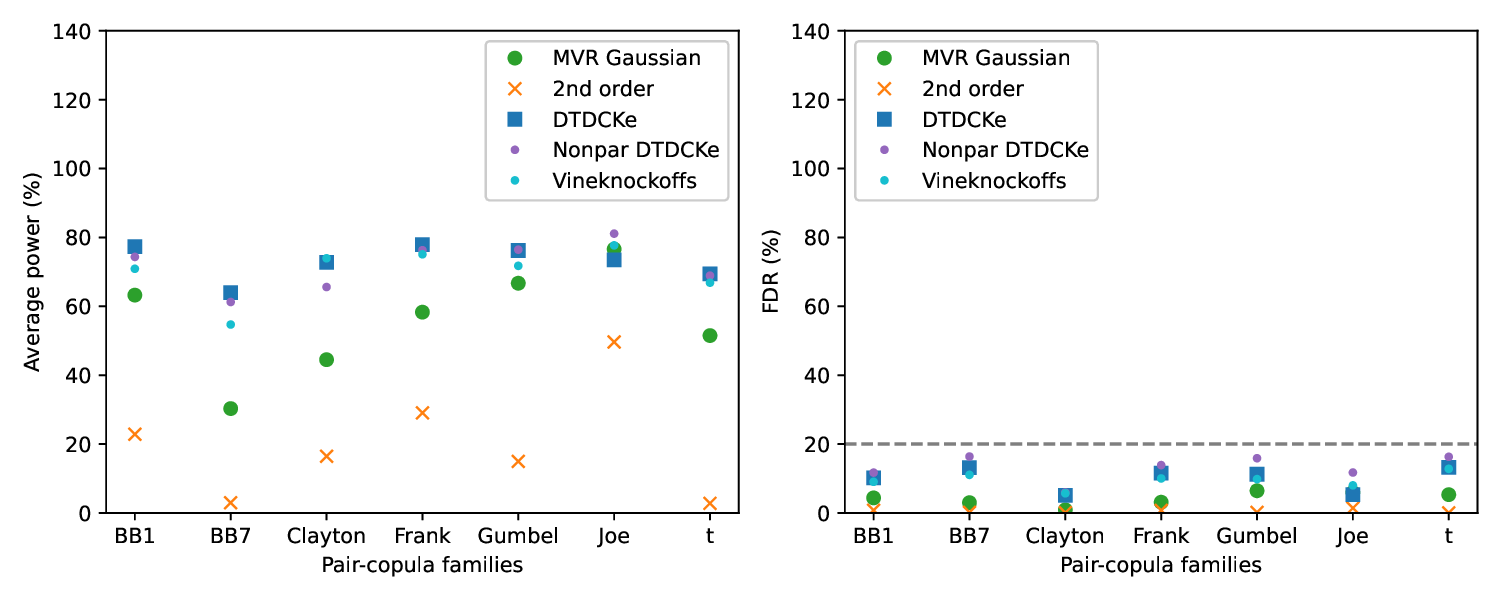}
    \caption{Empirical power and FDR for various copula families within the Truncated D-vine copulas DGP for a logistic linear response. Each data point in the graphs represents the average value obtained from 100 repetitions.}
    \label{Fig8}
\end{figure}

Figure \ref{Fig8} presents the power and the FDR for the logistic regression response across the different bivariate copula families. The parametric DTDCKe procedure demonstrates the highest statistical power for the BB1, BB7, Frank, Gumbel, and t copula families, closely followed by the non-parametric DTDCKe method. The Vineknockoff method exhibits the highest power for the Clayton copula family, followed closely by the parametric DTDCKe procedure. In the case of the Joe copula, the non-parametric DTDCKe method exhibits the best power performance. As before, the second-order knockoffs display the lowest statistical power. Similar to the Gaussian response, all methods effectively control the FDR at the predetermined level. 

\begin{remark}
    Results in Appendix \ref{sec:Multivariate Normal Distribution} indicate that, when measuring the false discovery proportion and the empirical power, the DTDCKe algorithms exhibit good performance for the DGP corresponding to a multivariate normal distribution for scenarios of high correlation. Furthermore, the results in Appendix \ref{sec:Gene D-vine} indicate that our proposals also exhibit good performance for the DGP corresponding to a parametric D-vine copula fitted to gene expression data; in this case, the good performance is maintained as the number of covariables increases.   
\end{remark}

\section{Application to gene expression data}
\label{sec:Application}

In this section, we apply the proposed methodologies to analyze a gene expression dataset to validate their effectiveness in selecting relevant variables. We perform a broad analysis using a publicly accessible dataset of gene expression data of lung tumors, obtained from the Lung Cancer Explorer (LCE) database \url{http://lce.biohpc.swmed.edu/}. The LCE database contains comprehensive lung cancer-specific expression data and clinical and survival information from 56 studies. These valuable data were collected from 23 different repositories, undergoing meticulous processing, standardization, and rigorous curation \citep{cai2019lce}.

We analyze the \verb|TCGA\_LUAD\_2016| genomic dataset introduced originally by the \cite{cancer2014comprehensive}. We aim to identify genes contributing to the differentiation between normal and lung cancer tissue. The considered dataset contains the gene profile of 576 patients, including 517 lung adenocarcinoma tumor samples and 59 normal tissues. Specifically, this dataset includes a vast collection of 20,419 gene expressions alongside pertinent clinical information, such as gender, age, race, smoking status, and tumor stage.

The \verb|TCGA\_LUAD\_2016| dataset presents a substantial class imbalance, with only 10.24\% of the profiled tissues belonging to the normal class (no cancer). We address this issue by applying the Synthetic Minority Over-sampling Technique (SMOTE) to generate synthetic minority instances \citep{chawla2002smote}. Despite its widespread use, the SMOTE introduces additional randomness, resulting in different oversampled datasets each time it is applied. Such variability may impact the reliability of the variable selection methods. Thus, we adopt a stable variant known as the Stable SMOTE to enhance replicability \citep{feng2021investigation}. This approach reduces the randomness at each step of the SMOTE process, producing more stable balanced datasets and contributing to a more consistent variable selection in scenarios with highly unbalanced binary responses.

Previous investigations have shown that applying a variance filter before analyzing gene expression datasets tends to increase the discovery of relevant genes \citep{hackstadt2009filtering,bourgon2010independent}. Also, our simulations demonstrate that the DTDCKe procedures exhibit a decrease in statistical power when the ratio $p/n$ increases (see Section \ref{sec:Gene D-vine} of Appendix \ref{sec:Simulations}). Consequently, we incorporate a variance prefiltering step in our analysis of the \verb|TCGA\_LUAD\_2016| dataset to improve the statistical power of the DTDCKe procedures.

To start our analysis of the \verb|TCGA\_LUAD\_2016| dataset, we oversample the minority class and generate 458 new minority instances using the Stable SMOTE technique, resulting in an augmented dataset of sample size $n= 1034$. Our numerical simulations for gene expressions with logistic responses described in Section \ref{sec:Gene D-vine} indicate that, with a ratio $p/n=0.25$, the DTDCKe methods exhibit a high statistical power of around 90\%. Therefore, motivated by these experimental findings, we select the 250 most variable genes to create the final dataset, achieving an approximate ratio of $p/n=0.25$.

We apply the parametric and non-parametric DTDCKe procedures with a target FDR control of $\alpha_{ebh}=0.20$. Additionally, we derandomize the Vineknockoff, MVR Gaussian, and second-order filters employing the DKe method. The lowest levels for the FDR control that allow for finding a data-dependent threshold in these last three methods are $\alpha_{ebh}=0.20$, $\alpha_{ebh}=0.56$, and $\alpha_{ebh}=0.57$, respectively; therefore, we decided to employ these control levels with the described knockoff filters. The features selected by each of the five methods considered are presented in Table \ref{tab:1}.

To assess the validity of our findings, we compare our selections with the results of the tumor vs. normal meta-analysis of the LCE web application \citep{cai2019lce}. In the LCE tumor vs. normal meta-analysis module, each gene is associated with an estimate and a 95\% confidence interval (CI) for the standardized mean difference (SMD). The meta-analysis module also indicates the corresponding p-value for testing the overall effect. The meta-analysis accounts for heterogeneity using a random-effects model, considering only studies that include the analyzed genes \citep{cai2019lce}. Table \ref{tab:1} also presents this additional information to provide a comprehensive analysis.

Table \ref{tab:1} illustrates that the non-parametric DTDKCe procedure identifies the highest number of variables (23) and that the parametric DTDKCe and Vineknockoffs procedures identify 16 and 11 variables, respectively. Contrastingly, the performance of the MVR Gaussian knockoff filter is deficient compared to the simulation results presented in Section \ref{sec:Gene D-vine} of Appendix \ref{sec:Simulations}. With a target level of $\alpha_{ebh}=0.56$ for the FDR, the filter selects only six features. Similarly, the second-order knockoff filter underperforms relative to the simulations in Appendix \ref{sec:Simulations}, identifying only seven features with a target level of $\alpha_{ebh}=0.57$ for the FDR.

\begin{table}[ht]
    \caption{Gene features identified by the knockoff filters. The table also includes information on the tumor vs. normal meta-analysis from the LCE web application: an estimate of the pooled standardized mean difference (SMD) with a 95\% confidence interval (CI) and the p-value for testing the overall effect.}
    \label{tab:1}
    
    \centering

   \begin{tabular}{cm{0.8cm}m{0.8cm}m{0.8cm}m{0.8cm}m{0.8cm}cc}
        \hline
         & \multicolumn{1}{l}{\rot{DTDCKe} \rot{FDR = 0.20}} & \multicolumn{1}{l}{\rot{Non DTDCKe} \rot{FDR = 0.20}} & \multicolumn{1}{l}{\rot{Gaussian} \rot{FDR = 0.56}} & \multicolumn{1}{l}{\rot{2nd-order} \rot{FDR = 0.57}} & \multicolumn{1}{l}{\rot{Vineknockoff} \rot{FDR = 0.20}} &  \multicolumn{2}{c}{Meta-analysis}\\
         Gene &  &  &  &  &  & SMD [95\% CI] & P-Value\\
         \hline\centering
         INHA &    & \checkmark &     &  &  & $0.75$ $[0.57; 0.93]$ & $p<0.01^{**}$\\
         B4GALNT4 &  \checkmark  &  \checkmark   &   &  & \checkmark  & $1.42$ $[1.04, 1.80]$  & $p<0.01^{**}$ \\
         AKR1B10 &  \checkmark  & \checkmark  &   &  &  & $0.93$ $[0.67, 1.18]$ & $p<0.01^{**}$ \\
         PAX7 &   & \checkmark   &  &  &  & $0.57$ $[ 0.40, 0.75]$ & $p<0.01^{**}$ \\
         CYP24A1 &  \checkmark & \checkmark  &   &  & \checkmark & $1.29$ $[0.87, 1.70]$ & $p<0.01^{**}$ \\
         HTR3A & \checkmark  & \checkmark  &   &   & \checkmark & $1.23$ $[ 0.82, 1.63]$ & $p<0.01^{**}$ \\
         SYT12 &  \checkmark & \checkmark  &   & \checkmark   &  \checkmark & $1.30$ $[ 0.76, 1.84]$ & $p<0.01^{**}$ \\
         FAM83A  & \checkmark & \checkmark  &   &  &  \checkmark  & $2.16$ $[1.37, 2.95]$ & $p<0.01^{**}$ \\
         ABCA12 & \checkmark & \checkmark &   &  &  \checkmark & $1.23$ $[0.82, 1.65]$ & $p<0.01^{**}$ \\
         HORMAD1 &  & \checkmark  &  &   &  & $1.03$ $[0.66, 1.40]$ & $p<0.01^{**}$ \\
         C20orf85 &  \checkmark &  \checkmark  & &  &  & $-1.12$ $[-1.54, -0.71]$ & $p<0.01^{**}$ \\
         AKR7A3  &\checkmark & \checkmark &  &   &  \checkmark & $1.00$ $[0.62, 1.37]$ & $p<0.01^{**}$ \\
         AFAP1-AS1  & \checkmark & \checkmark & \checkmark  & \checkmark  &  \checkmark & $2.19$ $[1.33, 3.05]$ & $p<0.01^{**}$ \\
         HOXC13 & \checkmark  & \checkmark &   &   &  & $1.01$ $[ 0.60, 1.42]$ & $p<0.01^{**}$\\
         ITLN2 & \checkmark & \checkmark & \checkmark & \checkmark  &  \checkmark & $-3.07$ $[-4.52, -1.62]$ & $p<0.01^{**}$\\
         MUC6 & \checkmark  & \checkmark & \checkmark  & \checkmark  &  & $0.43$ $[ 0.20, 0.65]$ & $p<0.01^{**}$ \\
         C8B & \checkmark  & \checkmark & \checkmark  & \checkmark  &  \checkmark & $-1.02$ $[-1.61, -0.43]$ & $p<0.01^{**}$ \\
         MAGEA12 &   & \checkmark &   &  &  & $0.64$ $[0.26, 1.01]$ & $p<0.01^{**}$ \\
         GLB1L3 & \checkmark  & \checkmark & \checkmark  & \checkmark  &  \checkmark & $0.63$ $[ 0.13, 1.12]$ & $p=0.013^{*}$ \\
         ALOX15 &   & \checkmark &  &   &  & $-0.46$ $[-0.92, -0.01]$ & $p=0.046^{*}$ \\
         PCSK2 &   & \checkmark &  &   &  & $-0.37$ $[-0.77, 0.02]$ & $p=0.064$ \\
         GSTM1 &   & \checkmark &  &   &  & $-0.28$ $[-0.92, 0.36]$ & $p=0.39$ \\
         APOH & \checkmark  & \checkmark & \checkmark & \checkmark  &  & $-0.14$ $[-0.84, 0.56]$ & $p=0.69$ \\
          \hline
         \\
    \end{tabular}
\end{table}


Considering the meta-analysis of statistically significant genetic features ($p < 0.05$), the false discovery proportion (FDP) is calculated by dividing the number of non-statistically significant genetic features by the total number of selected genes. The FDP values are 6.3\% for the parametric DTDCKe, 13.0\% for the non-parametric DTDCKe, 0\% for the Vineknockoffs, 16.7\% for the MVR Gaussian, and 14.3\% for the second-order knockoff filter. All five knockoff filters achieve FDPs below the pre-specified level. Consequently, the non-parametric DTDCKe method is the most effective knockoff filter, identifying the highest number of non-null variables as 19. It is followed by the parametric DTDCKe and Vineknockoffs procedures, which identify 14 and 11 relevant features, respectively. In contrast, the MVR Gaussian and second-order knockoff filters perform the worst, selecting five and six non-null variables, respectively. It is important to note that we cannot estimate the statistical power in this analysis since the genes genuinely related to the disease are unknown.

In many real-world exploratory analyses, the goal is usually to detect as many genes as possible that are genuinely associated with a phenotype. The results above show that the DTDCKe procedures provide a helpful new tool for this task. Consequently, we recommend applying the non-parametric DTDCKe along with its parametric version or the Vineknockoffs method to strengthen the selection and its reliability.

\section{Discussion}
\label{sec:Discussion}

This paper introduces the Derandomized Truncated D-vine Copula
Knockoffs with e-values (DTDCKe) procedure, which addresses the
challenge of sampling knockoffs in non-Gaussian environments in the presence of complex joint dependence structures. Using truncated D-vine copulas to model the
joint distribution of the vector $(\bm{X}, \tilde{\bm{X}})$ effectively diminishes dependency between each
variable $X_j$ and its corresponding knockoff $\tilde{X}_j$. This
characteristic significantly enhances the statistical power of the
proposed method, especially when dealing with highly correlated
predictors. Notably, wrapping the TDCK algorithm with the Derandomized
Knockoffs with e-values algorithm proposed by
\cite{ren2022derandomized} produces a more steady variable selection
method.

Through extensive simulation studies, we empirically show that the
DTDCKe procedure consistently controls the FDR across various
scenarios involving Gaussian and logistic linear responses. Compared
to the Vineknockoffs method, the DTDCKe procedures exhibit better
statistical power across nearly all considered DGPs, while
Vineknockoffs enforces a more stringent FDR control. Our proposals
perform much better than Vineknockoffs when the distribution of the
covariables possesses heavy tails; of all the cases considered in the
simulation study, Vineknockoffs seem to have better results only in
the highly correlated Gaussian scenario. Furthermore, the DTDCKe
methods demonstrate superior statistical power performance compared to
the MVR Gaussian and second-order knockoffs in all assessed DGPs.

To validate the usefulness of our approach in an actual situation, we
apply our methodology to a gene expression dataset of lung tumors. The
primary objective of the study was to identify genes
contributing to the differentiation between normal and lung cancer
tissue. According to a univariate meta-analysis conducted using the
LCE web application, the proposed methods are more robust and accurate
in identifying relevant genes than the competing knockoff methods
considered here. Based on the analysis and the results of the study, it is fair to claim that the Truncated D-vine approach
has a remarkable potential in modeling gene expression data.

The DTDCKe procedure possesses several compelling features. One
remarkable feature of the methodology is that it allows for the straightforward use of the available tools in the \verb|R| package \verb|rvinecopulib|. This library provides a
framework for effective model-fitting using various parametric
pair-copula families, including rotated variants. The
\verb|rvinecopulib| package also implements the mBICV copula selection
criterion, which is highly suitable for high-dimensional
scenarios. Another significant advantage of using the
\verb|rvinecopulib| package is that it allows for fitting non-parametric bivariate copulas, providing a viable time-saving
alternative to the parametric DTDCKe procedure. Our simulations
demonstrate that the estimation time for the non-parametric DTDCKe is only 20\% of the time required by the parametric DTDCKe. This feature is especially valuable in high-dimensional settings
despite the slight inflation of the FDR displayed by the non-parametric DTDCKe procedure.

Another advantage of the proposed procedure is its ability to employ non-parametric marginal distribution transformations. This approach consists of pulling the observation into the copula scale
using pseudo-observations, generating the knockoff variables in this
scale, and reversing the transformation using the quantile function, eventually obtaining
knockoffs in the scale of the original marginal
distributions. Such transformations eliminate the need to select a
specific parametric family or use a kernel density estimator to
specify the marginal distributions, simplifying the estimation
process. Consequently, they provide a remarkably versatile and broadly
applicable knockoff framework.

Additionally, it is crucial to underscore the substantial benefits of leveraging parallel computation with the \verb|rvinecopulib| package. When applied to the gene expression data, which comprises a matrix with 1034 rows and 250 columns after applying the Stable SMOTE and variance filter preprocessing, the parametric and non-parametric DTDCKe procedures require approximately 25.3 minutes and 14.5 minutes, respectively, utilizing 23 cores. These runtimes represent a mere 16\% and 8\% of the time the Vineknockoffs method needs, which takes around 153 minutes due to its inability to parallelize the knockoff generation process. Thanks to all the advantages we enlisted for the DTDCKe procedure and the rich options of the \verb|rvinecopulib| package, implementing the DTDCKe procedure is straightforward. All the source codes and the datasets necessary for
reproducing all analyses in this paper are publicly available at \url{https://github.com/AlejandroRomanVasquez/DTDCKe}.

Despite the proposed methodology's compelling properties, some unresolved questions remain. Firstly, the symmetry of the D-vine structure plays a crucial role in generating knockoffs that effectively control the FDR. Expanding the scope of this approach to include the broader context of the R-vine copulas for more precise modeling of the joint distribution of the features $\bm{X}$ offers a captivating path for future research. Secondly, the current DTDCKe procedure is limited to continuous variables. Therefore, extending vine copula modeling to include discrete variables or mixed discrete-continuous scenarios would undoubtedly enhance the applicability of the proposed knockoff sampling method. Lastly, while D-vine copulas offer flexibility in modeling dependency structures, this approach is built on simplifying assumptions that may not hold under certain circumstances; see, for instance, \cite{kurz2022testing}. A promising future investigation direction is exploring non-simplified models, including the generalized additive models for bivariate copula constructions \citep{vatter2018generalized}, and the non-parametric methods based on splines \citep{schellhase2018estimating}.

\noindent\textbf{Acknowledgements:} Partial support from CONAHCyT project CBF2023-2024-3976. Alejandro Román Vásquez acknowledges a grant from CONAHCyT Estancias Posdoctorales por México 2022 at Centro de Investigación en Matemáticas.

\bibliography{references}

\begin{appendix}

\section{Simulation studies}
\label{sec:Simulations}

As mentioned in Section \ref{sec:Simulations_paper}, this appendix
presents the simulation results for the DGPs given by a multivariate
normal distribution and a parametric D-vine copula fitted to gene
expression data. Each DGP is considered under both Gaussian and
logistic linear response models. The number of non-null features is set to 20\% of the $p$ covariates, with the corresponding indices
randomly sampled for each simulated dataset. The non-null coefficients
are drawn from a uniform distribution $\textrm{Unif}\left(\left[
  -\delta, -\frac{\delta}{2}\right] \cup \left[ \frac{\delta}{2},
  \delta\right] \right)$ to emulate various positive and negative
effects. Following \cite{ren2022derandomized} and
\cite{romano2020deep}, we set $\delta = \alpha / \sqrt{n} $, where the
amplitude $\alpha$ takes different values depending on the DGP and the
regression response.

In the following, we recall some aspects of the configuration employed in the simulation studies of Section \ref{sec:Simulations_paper}, which we also use below. We simulate $n_{sim} = 100$ datasets for the regression models in each DGP. Our target FDR level is set to 20\%, and then we set $\alpha_{kn}=0.1$ and $\alpha_{ebh}=0.2$. We employ 5-fold cross-validation to optimize the regularization parameter $\lambda$ and fix $M_{lasso}=10$; in this manner, we try to balance computational efficiency and result robustness. We sample $M=50$ knockoffs when applying the DKe algorithm. The following three subsections provide further details regarding other parameters of the simulation experiments for each DGP. The results of the simulations are shown in the corresponding line plots, which display the relationship between the average power and the FDR against a changing parameter.

\subsection{Multivariate Normal Distribution}
\label{sec:Multivariate Normal Distribution}

The fundamental Gaussian case serves as an appealing instance for which our proposals should perform well. We consider a multivariate normal distribution $\bm{X} \sim N_p\left(
\textbf{0}, \bm{\Sigma} \right)$ with exponentially decaying
correlations given by $\bm{\Sigma}_{ij}= \rho^{|i-j|}$, $\rho>0$. In
this case, we set the sample size at $n=300$, the total number of
features at $p=100$, and the amplitude at $\alpha=10$. Our proposal
might help in scenarios of high correlation since, as previously
stated, one of the objectives of constructing a $p$-truncated
$2p$-dimensional D-vine copula for $(\bm{X},\tilde{\bm{X}})$ is to
diminish the dependence between $\bm{X}$ and $\tilde{\bm{X}}$. Hence,
it is preponderant to assess the impact of varying the correlation
coefficients in high correlation scenarios,  which range from 0.7 to 0.9.

\begin{figure}[ht]
    \centering
    \includegraphics[width=1\textwidth]{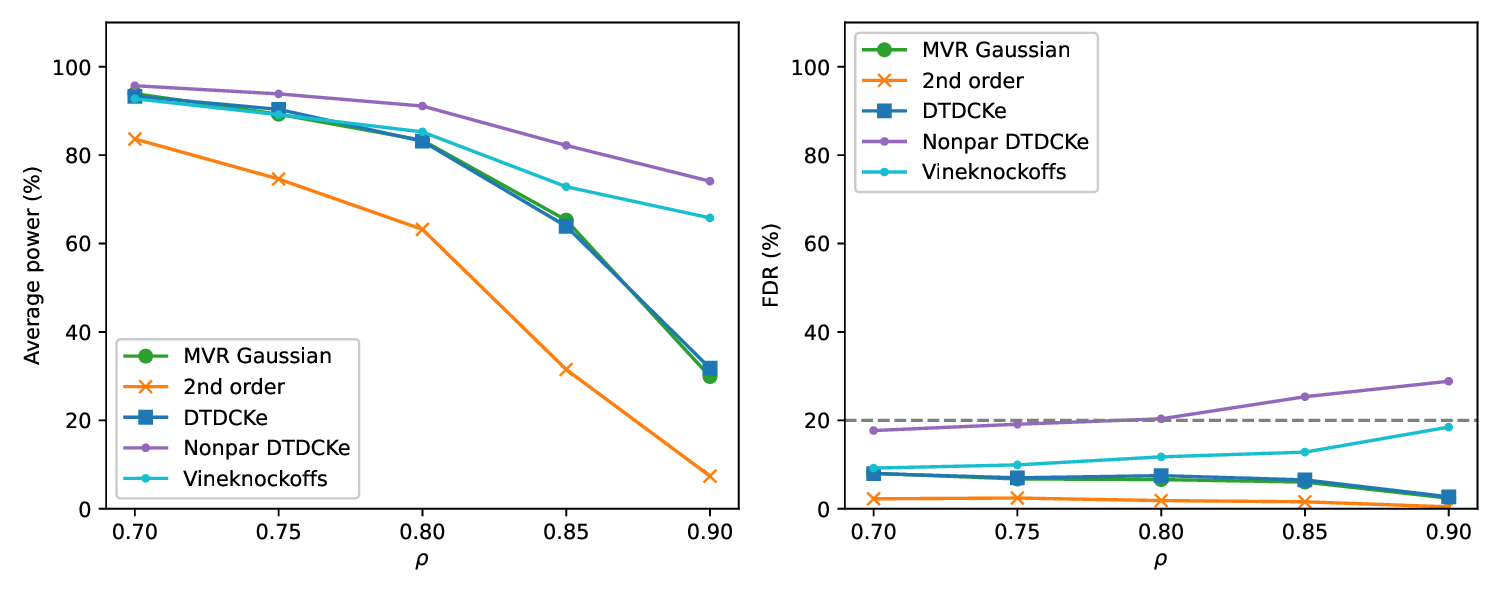}
    \caption{ Empirical power and FDR as $\rho$ varies in the multivariate normal distribution DGP for a Gaussian linear response. Each point in the graphs represents the average value across 100 repetitions.}
    \label{Fig2}
\end{figure}

Figure \ref{Fig2} presents the simulation experiments' empirical power and FDR for the Gaussian response. The performance of the parametric DTDCKe procedure is highly similar to that of the MVR Gaussian knockoffs method. Both approaches effectively control the FDR at comparable levels and exhibit similar power, although the parametric DTDCKe procedure shows slightly higher power when $\rho>0.85$. In contrast, the non-parametric DTDCKe procedure has better power across the range of $\rho$ values; however, it displays a slightly elevated FDR control, which becomes more pronounced for $\rho>0.80$. The Vineknockoffs method demonstrates a balanced performance throughout, exhibiting power metrics greater than those of the parametric DTDCKe but lower than those of the non-parametric DTDCKe procedure. Additionally, it maintains effective FDR control across all values of $\rho$. Laslty, despite maintaining a good FDR control, the second-order knockoffs consistently exhibit lower power than the other knockoff filters, with a significant decrease when $\rho>0.80$.

\begin{figure}[ht]
    \centering
    \includegraphics[width=1\textwidth]{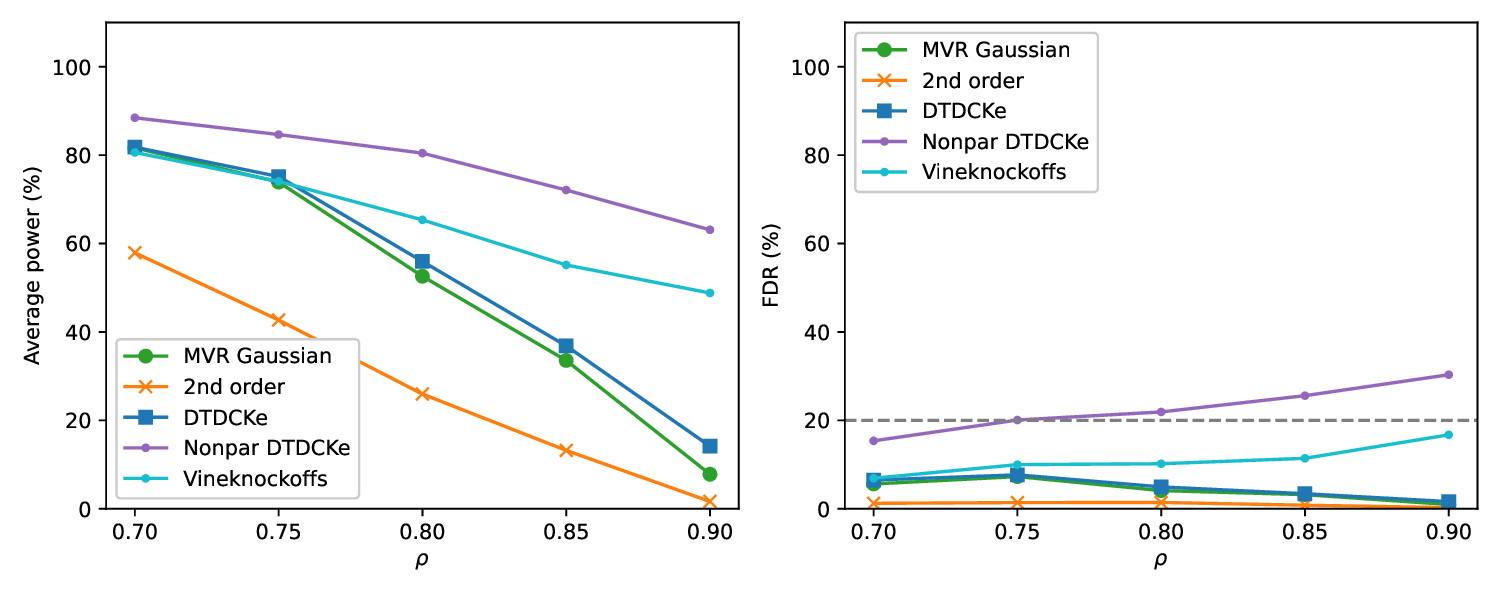}
    \caption{Empirical power and FDR as $\rho$ varies in the multivariate normal distribution DGP for a logistic linear response. Each point in the graphs represents the average value across 100 repetitions.}
    \label{Fig6}
\end{figure}

Figure \ref{Fig6} shows the simulation results for the logistic response. The configuration setups for the Gaussian linear response are retained, except for the amplitude $\alpha$, which is set to 70 to have an appropriate statistical power. Statistical power and FDR trends are similar to those observed in the Gaussian regression. Both the parametric DTDCKe procedure and the MVR Gaussian knockoffs exhibit comparable results regarding the statistical power and the FDR control. However, the parametric DTDCKe procedure shows slightly superior statistical power, particularly when $\rho>0.70$. Also, the non-parametric DTDCKe procedure appears to exhibit the highest power performance but fails to maintain proper control of the FDR when $\rho>0.80$. The Vineknockoffs method achieves power levels between the parametric DTDCKe and non-parametric DTDCKe procedures, controlling the FDR across all values of $\rho$. The second-order knockoff filter shows the poorest performance in terms of power.

\subsection{ D-vine copula fitted to gene expression data }
\label{sec:Gene D-vine}

In this subsection, we aim to emulate the multivariate probability
structure that is inherent in real gene expression datasets, such as the one studied in the illustration of Section \ref{sec:Application}, thereby serving
as a DGP for the covariables $\bm{X}$. Motivated by using vine copula
models to construct graphical gene networks
\citep{chatrabgoun2020constructing, farnoudkia2021vine}, we employ a
D-vine copula to model a concrete genomic dataset. This approach
enables us to capture the complex joint dependence structures that are likely to be present in this type of data.
The resulting fitted vine copula object
is subsequently employed to simulate predictors $\bm{X}$. We employ
the real-world genomic dataset outlined in
\cite{rousseaux2013ectopic}, which encompasses 307 tumor samples from
lung cancer patients, comprising an extensive set of 20,356 genes and
additional clinical information, including tumor stage, age, and gender. This dataset is publicly accessible through the Lung
Cancer Explorer (LCE) web portal \citep{cai2019lce}.

A widely adopted approach to enhance gene-disease discoveries in gene expression studies involves preselecting the most highly variable genes \citep{bommert2022benchmark}. Thus, in order to examine the influence of various variance thresholds, we consider five distinct subsets of the genomic dataset described above. Each subset is determined after applying a variance filter; hence, they include the most variable genes for their corresponding size. After applying the variance filter, the number of retained covariables in these subsets is set to $p=100, 125, 150, 175, 200$. We then proceed to fit a parametric D-vine copula model to each of these subsets. 

We adapt the steps described in Section
\ref{sec:Knockoffs_vine_copulas} to fit the parametric D-vine
copulas. The main differences with the procedure described in Section
\ref{sec:Knockoffs_vine_copulas} are: 1) we fit a complete D-vine to
the data without truncation; 2) we do not include the knockoff
variables yet; and 3) we only consider parametric families. We start
determining the order of the variables to fit the parametric D-vine
copula models. Then, we transform the gene expression data to the
copula scale. Next, we fit the copula model in the copula scale,
basing the selection of the family on the mBICV criterion and
estimating the parameters with maximum likelihood. Using the fitted
D-vine copula, we simulate $n=400$ observations of the copula,
representing realizations of each variable on the copula scale. We
revert to the original individual gene scale employing the empirical
quantile function described in Section
\ref{sec:Knockoffs_vine_copulas}. Thus, we generate synthetic data by approximating the dependence structure of the considered gene
expression dataset. We use this synthetic data to simulate the
Gaussian and logistic responses required to complete the simulation
study.

\begin{figure}[ht]
    \centering
    \includegraphics[width=1\textwidth]{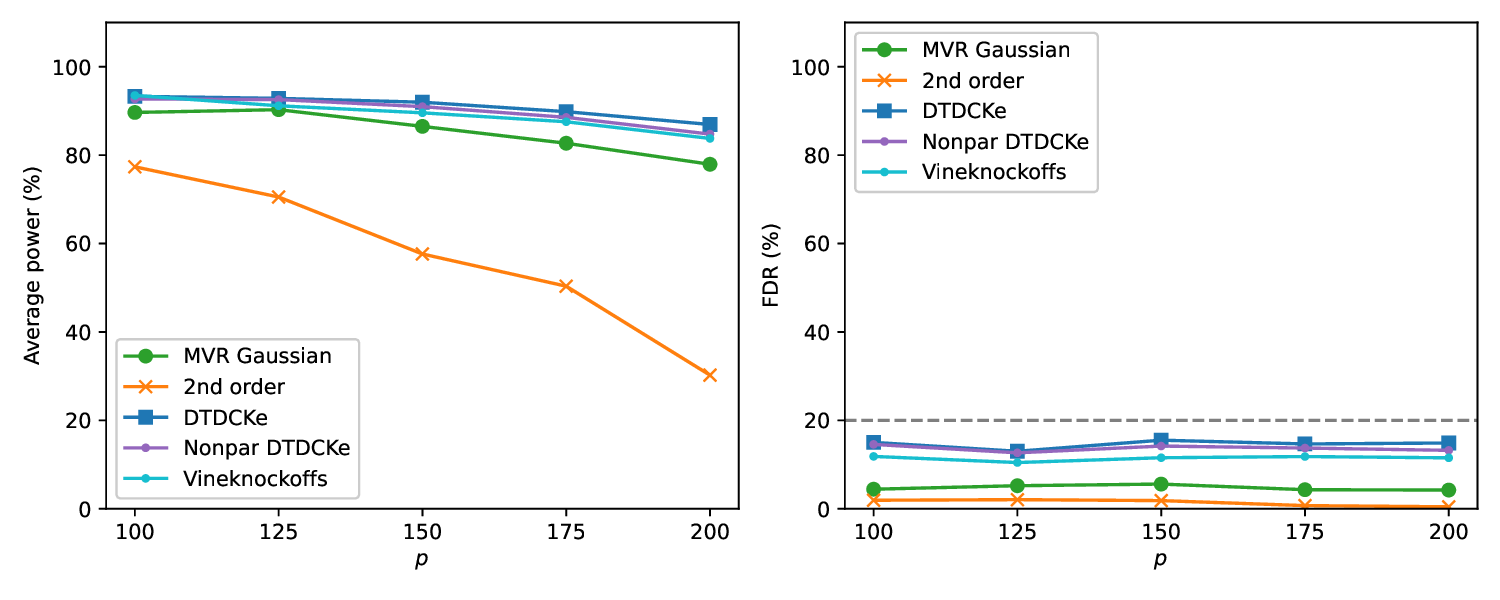}
    \caption{Empirical power and FDR as a function of the number of covariables $p$ for the DGP associated with a parametric D-vine copula model fitted to gene expression data and a Gaussian linear response. Each data point on the graphs represents the average value of 100 repetitions.}
    \label{Fig5}
\end{figure}

Figure \ref{Fig5} illustrates the main findings for the Gaussian linear response, with an amplitude of $\alpha=8$. As the number of features $p$ increases (i.e., the $p/n$ ratio grows), the statistical power of the five considered methods decreases. Both DTDCKe procedures exhibit remarkably similar good power performances, with power values around 93\% for $p=100$ and dropping to approximately 85\% for $p=200$. The Vineknockoff method demonstrates analogous power metrics to DTDCKe procedures, albeit with slightly lower values. The MVR Gaussian knockoffs exhibit less statistical power than the previous procedures, but their performance is satisfactory, ranging from around 89\% for $p=100$ to nearly 78\% for $p=200$. In addition, the MVR Gaussian knockoffs’ power decreases slightly more than that of the proposed methods when the $p/n$ ratio increases. In contrast, the second-order knockoffs display the poorest performance, significantly affected as the number of variables increases. All methods effectively control the FDR at the predetermined level of 20\%.

For the logistic linear response, maintaining the configuration used
for the Gaussian response results in a dataset  that exhibits
significant
class imbalance. This problem is also detected in the D-vine
simulation study of Section \ref{sec:TDcopula_sim}. Therefore, to
achieve a more balanced class distribution, we modify the simulation
configuration for the logistic response. We employ the same method
used to generate the logistic regression in Section
\ref{sec:TDcopula_sim}. Nevertheless, this approach alone does not
solve the imbalance problem
since it is still possible to simulate
datasets where the class imbalance problem persists. This
issue can
be primarily attributed to the diverse characteristics of the marginal
distributions of the expressed genes in the genomic dataset used to
fit the DGP. In the literature, it has been observed that some genes
exhibit symmetric distributions,
while others display characteristics
such as asymmetry, heavy tails, or bimodality in their expression
profiles \citep{chatrabgoun2020constructing}. Thus, the imbalance
problem is entirely solved by conducting multiple simulations and only
retaining those exhibiting a proportion of 1s ranging from 30\% to
70\%. This choosing process ensures the generation of more balanced
datasets. The only other modification in the configuration for the
simulation of the logistic response is that the amplitude parameter
$\alpha$ is set to $\alpha=60$ to guarantee appropriate statistical
power.

\begin{figure}[ht]
    \centering
    \includegraphics[width=1\textwidth]{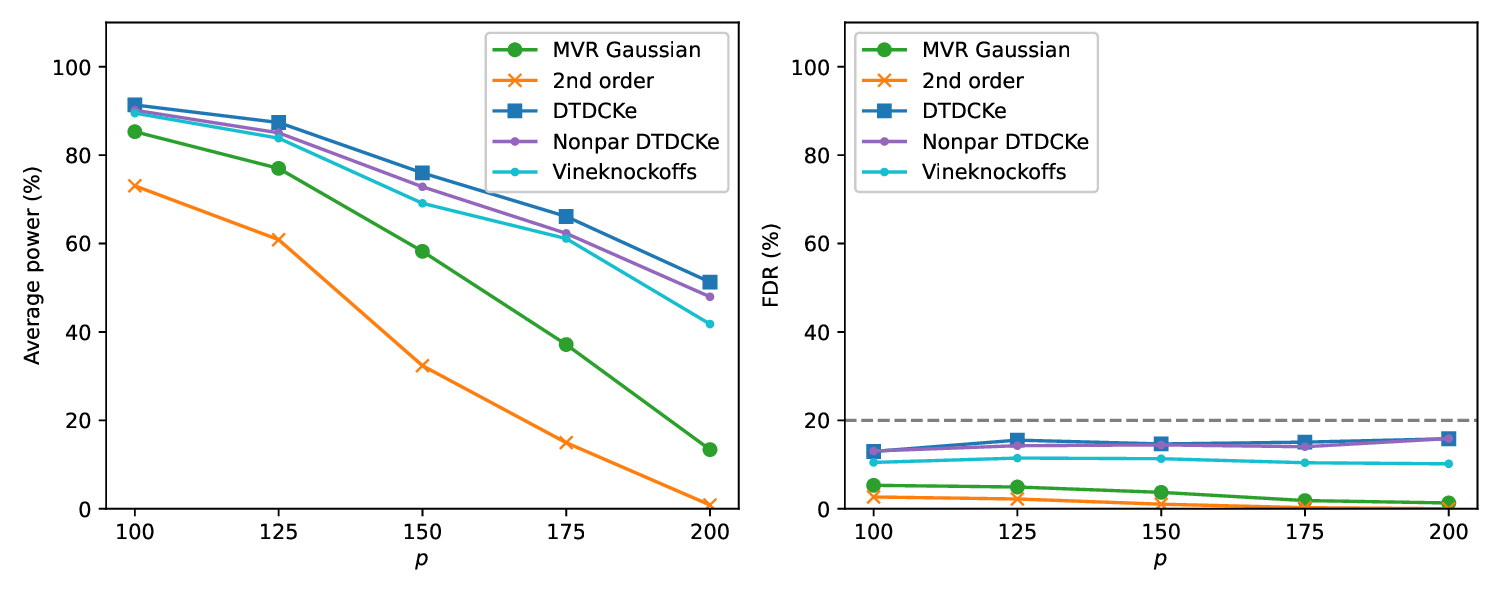}
    \caption{Empirical power and FDR as a function of the number of covariables $p$ for the DGP associated with a parametric D-vine copula model fitted to gene expression data and a logistic linear response. Each data point on the graphs represents the average value derived from 100 repetitions.}
    \label{Fig9}
\end{figure}

Figure \ref{Fig9} depicts the logistic linear response's empirical power and the FDR. Overall, the trends and behaviors closely resemble those observed for the Gaussian response. However, with the logistic response, the reduction in statistical power for all knockoff filters becomes more pronounced as the number of variables increases. Moreover, the disparity in statistical power between the DTDCKe procedures and the competitive methods (Vineknockoffs and MVR Gaussian) becomes more marked.

\end{appendix}

\end{document}